\documentclass[sigconf,nonacm, authorversion]{acmart}

\newcommand{\anoncite}[1]{[Reference Hidden]}
\usepackage{colortbl}

\usepackage{tabularx}

\usepackage{listings}

\lstdefinelanguage{prompt}{
    basicstyle=\ttfamily\scriptsize,
    breaklines=true,
    captionpos=b,
    frame=single, 
    rulecolor=\color{gray}, 
    frameround=tttt 
}

\AtBeginDocument{%
  }

\setcopyright{acmlicensed}
\copyrightyear{2025}
\acmYear{2025}
\acmDOI{XXXXXXX.XXXXXXX}

\acmConference[EASE 2025]{The 29th International Conference on Evaluation and Assessment in Software Engineering}{17–20 June, 2025}{Istanbul, Türkiye}
\acmISBN{978-1-4503-XXXX-X/18/06}
\setcopyright{none}

\begin{document}

\title{Quality Assessment of Python Tests Generated by Large Language Models}

\author{Victor Anthony Alves}
\email{anthonyvictor90@gmail.com}
\affiliation{%
  \institution{Federal University of Ceará (UFC)}
  \city{Quixadá}
  \state{Ceará}
  \country{Brazil}
}

\author{Carla Bezerra}
\affiliation{%
  \institution{Federal University of Ceará (UFC)}
  \city{Quixadá}
  \state{Ceará}
  \country{Brazil}
}
\email{carlailane@ufc.br}

\author{Ivan Machado}
\affiliation{%
  \institution{Federal University of Bahia (UFBA)}
  \city{Salvador}
  \state{Bahia}
  \country{Brazil}
}
\email{ivan.machado@ufba.br}

\author{Larissa Rocha}
\affiliation{%
  \institution{University of the State of Bahia (UNEB)}
  \city{Salvador}
  \state{Bahia}
  \country{Brazil}
}
\email{lari.rsoares@gmail.com}

\author{Tassio Virgínio}
\affiliation{%
  \institution{Federal University of Bahia (UFBA)}
  \city{Salvador}
  \state{Bahia}
  \country{Brazil}
}
\email{tassio.virginio@ufba.br}

\author{Publio Silva}
\affiliation{%
  \institution{Federal University of Ceará (UFC)}
  \city{Quixadá}
  \state{Ceará}
  \country{Brazil}
}
\email{publio.blenilio@gmail.com}

\renewcommand{\shortauthors}{Alves et al.}

\begin{abstract}
The manual generation of test scripts is a time-intensive, costly, and error-prone process, indicating the value of automated solutions. Large Language Models (LLMs) have shown great promise in this domain, leveraging their extensive knowledge to produce test code more efficiently. 
This study investigates the quality of Python test code generated by three LLMs: GPT-4o, Amazon Q, and LLama 3.3. 
We evaluate
the structural reliability of test suites generated under two distinct prompt contexts: Text2Code (T2C) and Code2Code (C2C). Our analysis includes the identification of errors and test smells, with a focus on correlating these issues to inadequate design patterns.
Our findings reveal that most test suites generated by the LLMs contained at least one error or test smell. Assertion errors were the most common, comprising 64\% of all identified errors, while the test smell \textit{Lack of Cohesion of Test Cases} was the most frequently detected (41\%). 
Prompt context significantly influenced test quality; textual prompts with detailed instructions often yielded tests with fewer errors but a higher incidence of test smells. Among the evaluated LLMs, GPT-4o produced the fewest errors in both contexts (10\% in C2C and 6\% in T2C), whereas Amazon Q had the highest error rates (19\% in C2C and 28\% in T2C). For test smells, Amazon Q had fewer detections in the C2C context (9\%), while LLama 3.3 performed best in the T2C context (10\%).
Additionally, we observed a strong relationship between specific errors, such as assertion or indentation issues, and test case cohesion smells. These findings demonstrate opportunities for improving the quality of test generation by LLMs and highlight the need for future research to explore optimized generation scenarios and better prompt engineering strategies.
\end{abstract}

\ccsdesc[500]{Software and its engineering~Software testing and debugging}
\ccsdesc[300]{Artificial Intelligence~Large Language Models}
\ccsdesc[100]{General and reference~Experimental studies}

\keywords{Large Language Models, Python Test Code, Test Smells, Prompt Engineering}


\maketitle

\section{Introduction}

The development of test scripts is an essential aspect of software engineering, ensuring the reliability and correctness of software systems \cite{gonzales2017}. However, creating test scripts manually is a time-consuming, labor-intensive, and expensive process, often prone to human error \cite{CHUN2022IEEE}. Consequently, researchers have extensively explored methods for automating test generation \cite{shamshiri2018, tufano2021unit}, leading to significant advancements in the field. Among these advancements, Artificial Intelligence (AI) tools have emerged as valuable assets in supporting automated code generation. By utilizing techniques such as search-based optimization and feedback-driven learning, these tools have enhanced their ability to produce efficient and accurate outputs \cite{Hansson_Ellréus_2023}.

Recent studies in Large Language Models (LLMs) have introduced new possibilities for test code generation \cite{SCHAFER2023IEEE, HAJI2024ACM, yetistiren2023, siddiq2024using, Fakhoury:TSE:2024, mathews2024}.
Models such as GPT-4, Amazon Q, and LLama 3.3 have demonstrated their ability to leverage extensive pre-trained knowledge and contextual understanding to automate the generation of test scripts. These models can simulate various test scenarios, identify potential issues, and improve testing efficiency compared to traditional approaches \cite{shengcheng2023ieee}. 
However, the effectiveness of LLM-generated test scripts is influenced by several factors, including the context and structure of the input prompts \cite{HAJI2024ACM}.


Prompt engineering, the process of crafting effective prompts to guide model outputs, has proven to be a critical factor in optimizing the performance of LLMs \cite{sahoo2024systematicsurveypromptengineering, Marvin2024}. Well-designed prompts can significantly improve the quality, structure, and usability of generated test code \cite{sahoo2024systematicsurveypromptengineering, Cain2024}. Research has shown that LLMs can adapt to different types of prompts, such as textual descriptions (Text2Code) and pre-existing code snippets (Code2Code), producing varying levels of quality depending on the generation context \cite{agarwal2024structuredcoderepresentationsenable, codexglue}. This adaptability highlights the transformative potential of LLMs in automating software testing.

Despite their capabilities, LLM-generated test scripts often exhibit issues related to errors and test smells \cite{siddiq2024using}. Like production code, test code must adhere to proper programming practices to ensure reliability and maintainability \cite{pynose}. Test smells, which refer to suboptimal design or implementation choices in test code, can reduce the effectiveness of tests in detecting faults and validating software behavior \cite{tufano2016, tufano2021unit}. For instance, studies have identified significant numbers of test smells in LLM-generated tests, such as those produced by GitHub Copilot \cite{alves2024sbes}. Common problems include assertion errors, poor cohesion among test cases, and inadequate error handling, highlighting the need for further investigation into LLM performance across diverse contexts.

This study builds upon the work of \citet{alves2024sbes}, extending the analysis of LLM-generated test code quality across three prominent models: GPT-4o, Amazon Q, and LLama 3.3. We compare the performance of these models in two prompt contexts, Text2Code and Code2Code, to evaluate their ability to produce high-quality Python test scripts. Our research focuses on the structural reliability of the generated tests, analyzing errors and test smells to identify patterns associated with suboptimal design. By investigating the relationship between prompt contexts, errors, and test smells, we aim to uncover opportunities for improving LLM-based test generation and provide actionable insights for researchers and practitioners.

In addition to evaluating the performance of the models, this work emphasizes the role of prompt engineering as a mechanism for enhancing the usability and reliability of automated test generation. As LLMs continue to advance, understanding their limitations and optimizing their use in test engineering are critical steps toward improving software quality and reducing development costs.

\section{Background}

\subsection{Unit Testing \& Test Smells}
\label{subsec:unittest}

Unit testing aims to identify defects and verify the functionality of the smallest components of software \cite{graham2021foundations}. Automated frameworks like JUnit\footnote{\url{https://junit.org/junit5/}} have made this approach widely adopted, allowing for the frequent and automatic execution of unit test suites \cite{daka2014}. This practice is crucial to avoid programming mistakes and detect problems early in the development process \cite{PENG2021101347, khorikov2020unit}. 

In the realm of unit testing, specific issues known as test smells can significantly undermine the quality and effectiveness of test cases. 
These smells typically arise from suboptimal design decisions made during the creation of test cases, which can greatly reduce test effectiveness \cite{palomba2018, deursen2001}. Test smells often emerge when test code is first added to a repository and tend to persist, negatively impacting the software's maintainability and directly affecting its quality \cite{tufano2021unit, santana2020raide, kim2020}. The presence of test smells in test code can directly compromise internal quality attributes such as code cohesion, complexity and maintainability \cite{Damasceno_Bezerra_Campos_Machado_Coutinho_2023}. The literature presents various approaches to detecting test smells, ranging from natural language-based techniques for removing test smells \cite{manoel2024} to automated tools that analyze tests and provide quantitative data on the identified test smells \cite{pynose, tempy2022, pytest-smell2022, jhonatan2024}.

\subsection{Test Generation by LLMs}
\label{subsec:test geneation}

Automated test generation facilitates the production and execution of numerous inputs that thoroughly test software units \cite{XieN2006}. Traditional approaches to test generation have included model-driven, requirement-based, static analysis, and research-driven techniques \cite{mcminn2004, Maragathavalli2011SearchbasedST}.

LLMs have assumed a significant role in test generation driven by Natural Language Processing (NLP) \cite{shengcheng2023ieee}. The performance of the code generated shows little variation between the different language models, regardless of their size or the data set used for training \cite{tristan2024}. By training in human language input, LLMs such as GPT-4o\footnote{\url{https://openai.com/index/hello-gpt-4o/}} and LLama 3.3\footnote{\url{https://www.llama.com/}} can generate code snippets, documentation, and even repair bugs \cite{mathews2024}. This functionality is exemplified
in applications like GitHub Copilot and ChatGPT, which have been extensively studied for their capability in test code generation \cite{HAJI2024ACM, SCHAFER2023IEEE}. In addition, some frameworks and techniques have recently been studied to facilitate test generation by LLMs \cite{chen2024, wang2024, nadia2024}.

Despite these advancements, significant challenges remain in evaluating the quality of code generated by these models \cite{liu2024}. While they hold promise in automating repetitive coding tasks and enhancing productivity, understanding the reliability, correctness, and maintainability of the generated test code is critical. Addressing this gap is essential to unlocking the full potential of LLMs in automated software testing and ensuring their outputs align with best practices and quality standards. 

\subsection{Prompt Engineering}
\label{subsec:prompt engineering}

The process of prompt engineering involves strategically designing task-specific instructions, guiding the output of the model without changing the parameters \cite{sahoo2024systematicsurveypromptengineering}. 
Studies on response generation in prompting contexts
\cite{reeves2023, liu2024} highlight
that the effectiveness of prompts can vary significantly based on the specific context and objectives. 
These studies emphasize that the design of instructions is pivotal in producing accurate and reliable outputs.

Additionally, recent research on process automation with LLMs \cite{agarwal2024structuredcoderepresentationsenable, codexglue, ryan2024} has explored various types of prompt contexts that influence the outputs generated by LLMs. These studies have characterized different prompt strategies and their applications, shedding light on how carefully designed prompts can optimize the capabilities of LLMs to meet task-specific requirements. Among the commonly discussed prompt contexts are:

\begin{itemize}

\item \textbf{Code to Code (\textit{C2C}):} context in which the prompt is a code snippet and the output is also a code snippet. Used in production, maintenance, and code translation tasks. 

\item \textbf{Text to Code (\textit{T2C}):} context in which the prompt is a natural language text and the output is a code snippet. Used in code generation and refactoring tasks.

\item \textbf{Code to Text (\textit{C2T}):} context in which the prompt is a code snippet and the output is natural language text. Used in code summarization and documentation tasks.

\end{itemize}

\section{Related Work}

\citet{HAJI2024ACM} conducted an experiment with the Codex \cite{codex2021} version of GitHub Copilot, investigating the usability of different test cases generated using command prompts. They found that a comment combining instructive natural language with an example of code usage resulted in more usable test generations. Similarly, \citet{shengcheng2023ieee} analyzed ChatGPT's ability to generate mobile test scripts. Their findings indicate that ChatGPT can generate useful tests when provided with sufficient context and information about the project architecture. \citet{alves2024sbes} investigated the presence of test smells in OSS project tests generated by LLMs in a single generation and prompt context. Although the authors of the \citet{alves2024sbes} study carried out a code quality analysis, this analysis was limited to one LLM (Github Copilot with GPT-4) and one prompt. In this work, we used two prompt contexts and correlated the occurrence of errors in the code with test smells.

Several studies have already been conducted to assess the quality of the production code. \citet{yetistiren2022acm} examined the quality of the code produced by the GitHub Copilot. 
They found that while Copilot can generate syntactically valid code, many of the generated solutions still exhibit issues related to efficiency and design.
In complementary research, \citet{yetistiren2023} investigated the presence of code smells in codes generated by Amazon's CodeWhisperer, Copilot and ChatGPT. They discovered that certain code smells frequently reappear in the generated code and that the LLMs themselves can solve these maintenance issues. This recurring pattern of problems affecting code maintainability was also noted by \citet{Hansson_Ellréus_2023}, who highlighted that ChatGPT and GitHub Copilot have been more effective in producing quality code. Therefore, although numerous evaluations of the quality of code generated by LLMs exist, these analyses have primarily focused on production code and have not specifically examined test code. From these studies, we extracted code quality analysis techniques and adapted them for test code.

\section{Study Design}

\subsection{Goals and Research Questions}

This study aims to evaluate the quality of the test suites generated by LLMs in Python. The evaluation focuses on both the syntactic and semantic quality of the generated code, as well as the structure of the test cases, with particular attention to identifying potential design irregularities. 
Initially, we assessed the validity of the tests by identifying errors. Subsequently, we measured the structural quality using test smell metrics. These evaluations were conducted across multiple LLMs and in two distinct prompt contexts: C2C and T2C. To guide our analysis, we formulated four key research questions:

\begin{itemize}
    \item[\textbf{RQ$_1$:}] \textbf{\textit{What are the types and frequencies of errors reported in the test suites generated by LLMs during execution?}} This question seeks to identify the errors that occur during the execution of the test suites generated by the models in Python. Errors can include syntax problems, runtime exceptions or failures in the test logic. 

    \item[\textbf{RQ$_2$:}] \textbf{\textit{What is the distribution and what are the specific types of test smells identified in the test suites generated by LLMs?}} The second question investigates the presence of test smells in the code. The analysis includes identifying the most common types and their distribution among the different test suites.

    \item[\textbf{RQ$_3$:}] \textbf{\textit{How does the type of prompt influence the occurrence of errors and test smells in test suites generated by LLMs?}} In this question, the focus is on the impact of the type of prompt used (T2C or C2C) on the quality of the tests generated. The question assesses whether different input styles result in significant variations in the frequency of errors and test smells.

    \item[\textbf{RQ$_4$:}] \textbf{\textit{Is there a correlation between the presence of errors and the detection of test smells in the same test suite generated by LLMs?}} This question explores the correlation between the presence of execution errors and test smells in the same test suite. The aim is to see if there is a pattern in which structurally problematic tests (with test smells) are more likely to present errors. 
    
\end{itemize}

\subsection{Study Settings}

\begin{figure*}[t]
    \scriptsize
    \centering 
    \includegraphics[width=1\textwidth]{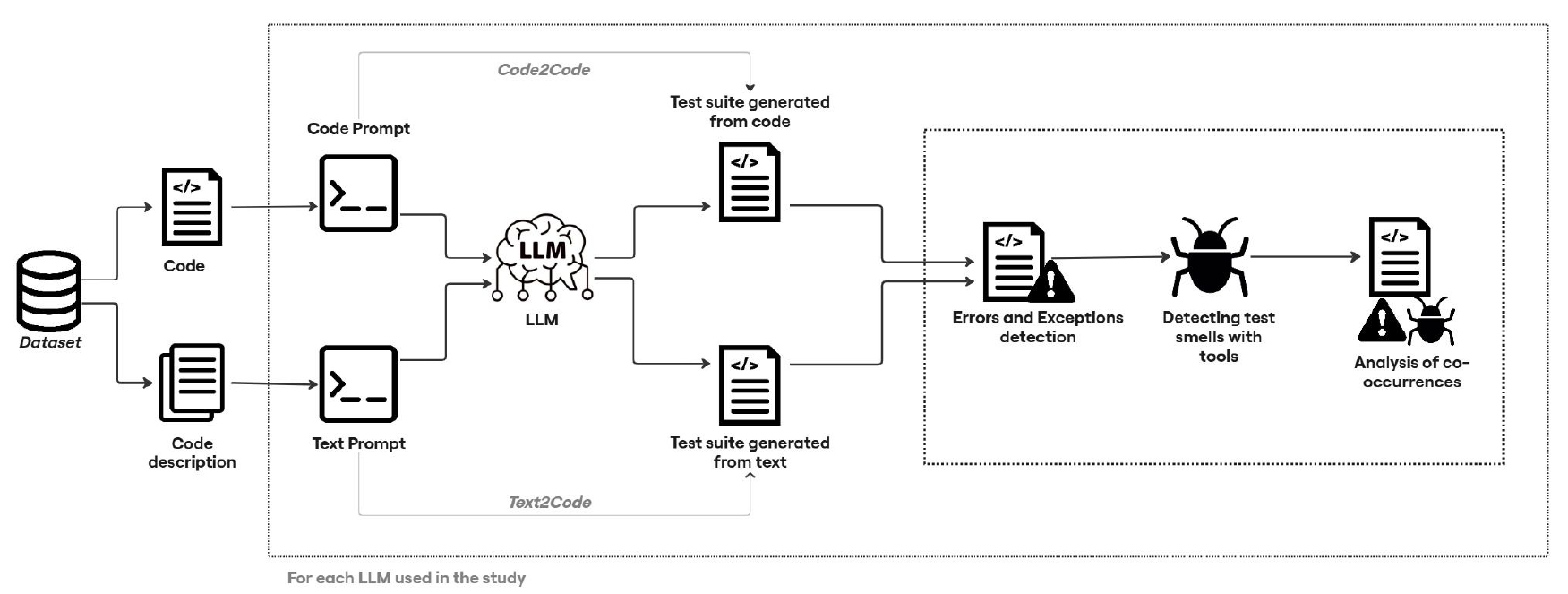}
    \caption{Methodological procedures.}
    \Description{A diagram outlining the methodological procedures, including dataset selection, prompt formulation, test suite generation, and analysis of errors and test smells.}
    \label{method}
\end{figure*}

We divided this study into four phases. First, we selected the dataset to be utilized. Next, we formulated the prompts. In the third phase, we generated the test suites using three LLMs. Finally, we identified errors, test smells, and co-occurrences to address the research questions. All the steps are outlined and mapped in Figure \ref{method}.

\subsubsection{Dataset Selection}

Initially, the codes and descriptions contained in the HumanEval\footnote{\url{https://github.com/openai/human-eval}} dataset were extracted and stored separately to facilitate the subsequent stages of the study. HumanEval, developed by OpenAI, was designed to train and evaluate LLMs' ability to generate functional code from natural language instructions and code examples. This dataset is made up of a collection of Python programming problems, each containing the final code, a textual description, input examples, and expected output.

\subsubsection{Preparing prompts}

After extracting the codes from the dataset and their respective textual descriptions, two different types of prompts were created in this study: one focused on code and the other on text.  The code prompt was composed predominantly of code extracted from HumanEval. Only one natural language instruction was added to guide test generation, ensuring the main focus remained on code analysis. In the text prompt, in addition to the textual description of the code provided by HumanEval, the prompt included a specific instruction to guide the generation of the tests; in this second context, the LLMs did not have access to the code, only to the text in natural language.

Listings \ref{lst:prompt1} and \ref{lst:prompt2} show examples of the prompts generated from the codes (C2C) and descriptions (T2C) contained in HumanEval. Listing \ref{lst:prompt1} shows an example of a textual prompt, including the name of the production code function, its description, and a specific instruction for generating the unit test. Listing \ref{lst:prompt2} displays a prompt containing only the instruction and the production code. It is worth noting that the code provided in this second prompt corresponds to the description used in the previous textual prompt, ensuring equivalence between the two input contexts.


\begin{lstlisting}[language=prompt, caption={Example of text prompt used in the study.}, label={lst:prompt1}]
from typing import List def separate_paren_groups(paren_string: str) -> List[str]:

The input to this function is a string containing several groups of nested parentheses. Its purpose is to separate these groups into separate strings and return a list of them. The separated groups are balanced (each open brace is closed correctly) and are not nested in each other. >>> separate_paren_groups('( ) (( )) (( )( ))') ['()', '(())', '(()())']

Write a unit test for this function according to the specifications
\end{lstlisting}

\begin{lstlisting}[language=prompt, caption={Example of code prompt used in the study.}, label={lst:prompt2}]
Write a unit test for this function:

from typing import List

def separate_paren_groups(paren_string: str) -> List[str]:
    result = []
    current_string = []
    current_depth = 0

    for c in paren_string:
        if c == '(':
            current_depth += 1
            current_string.append(c)
        elif c == ')':
            current_depth -= 1
            current_string.append(c)
            if current_depth == 0:
                result.append(''.join(current_string))
                current_string.clear()

    return result
\end{lstlisting}

\subsubsection{Test Code Generation}

The prompts prepared in the previous step were submitted to the LLMs so that they could generate automated test suites based on the information provided. Each item in the dataset contained a block of code accompanied by its description in natural language. Both were submitted separately to the LLMs so that, in the end, each code resulted in a pair of test suites: one generated based on the code (Code2Code context) and the other based on the textual description (Text2Code context). At the end of generation, these codes from the two prompt contexts were stored to be analyzed separately. For this study, three LLMs were used: LLaMA\footnote{\url{https://www.llama.com/}}, version 3.3 70B; Amazon Q Developer\footnote{\url{https://aws.amazon.com/pt/q/developer/}}; and GitHub Copilot\footnote{\url{https://github.com/features/copilot}}, version GPT-4o. 

\subsubsection{Analysis of Code Errors}

Once the test suites had been generated, they were run to check for interpretation errors. During this process, all errors reported by the Python interpreter were identified and mapped. To categorize the errors detected, the official Python documentation on Errors\footnote{\url{https://docs.python.org/3/tutorial/errors.html}} and Exceptions\footnote{\url{https://docs.python.org/3/library/exceptions.html}} was used as a reference, allowing for a detailed characterization of each error present in the generated suites. In this step, the distribution of error detection was counted and the error types were classified in relation to each prompt context, considering the results obtained for each of the three LLMs analyzed. To count the errors and exceptions, we initially added up all the detections identified in the overall set. Subsequently, we calculated the specific total for each type of error identified in each LLM, allowing for a segmented evaluation of the results, which is described in Equation \ref{eq:1}.

\begin{equation}\label{eq:1}
    T_{\text{errors}} = \sum_{i=1}^{m} \text{Error}_{i}
\end{equation}

In total, we found 151 detections of errors in the test suites. After counting all the errors, we calculated the percentage distribution of errors by type, as described in Equation \ref{eq:2}. The aim of this calculation is to determine the distribution of errors by type based on the total errors detected in each LLM (calculated in Equation \ref{eq:1}) in relation to the overall number of errors identified across all LLMs.

\begin{equation}\label{eq:2}
    Distrib_{\text{errors}} = \frac{T_{\text{errors}}}{\sum_{\text{LLMs}} T_{\text{errors}}}
\end{equation}

The errors detected in this study are categorized in Table \ref{erro-tipo}, which presents the error types along with their descriptions based on the Python language documentation. We identified eight main types of errors in the generated tests: \textit{AssertionError, SyntaxError, IndentationError, KeyError, AttributeError, NameError, TypeError}, and \textit{OverflowError.}

\begin{table}[H]
\small
\caption{Errors and Exceptions detected}
\label{erro-tipo}
\begin{tabular}{ll}
\rowcolor[HTML]{EFEFEF} 
\textbf{Error}   & \textbf{Description}                                                                                                                \\
AssertionError   & Raised when an assert statement fails.                                                                                              \\
\rowcolor[HTML]{EFEFEF} 
SyntaxError    & \begin{tabular}[c]{@{}l@{}}Raised when the code does not follow \\ the syntax rules of the Python language\end{tabular}             \\
IndentationError & \begin{tabular}[c]{@{}l@{}}Raised when the alignment of the lines \\ does not follow the blocks established\end{tabular}            \\
\rowcolor[HTML]{EFEFEF} 
KeyError         & \begin{tabular}[c]{@{}l@{}}Raised when a mapping (dictionary) \\ key is not found in the set of existing keys.\end{tabular}         \\
AttributeError   & \begin{tabular}[c]{@{}l@{}}Raised when an attribute reference or \\ assignment fails.\end{tabular}                                  \\
\rowcolor[HTML]{EFEFEF} 
NameError        & \begin{tabular}[c]{@{}l@{}}Raised when a local or global name is not \\ found. This applies only to unqualified names.\end{tabular} \\
TypeError        & \begin{tabular}[c]{@{}l@{}}Raised when an operation or function \\ is applied to an object of inappropriate type.\end{tabular}      \\
\rowcolor[HTML]{EFEFEF} 
OverflowError    & \begin{tabular}[c]{@{}l@{}}Raised when the result of an arithmetic \\ operation is too large to be represented.\end{tabular}       
\end{tabular}
\end{table}

\subsubsection{Test Smells Detection}

Once the test suites had been generated and the errors identified, the detection of test smells in the code was carried out. For this step, we previously selected two specific tools to identify inconsistencies and quality violations in unit tests: Pynose \cite{pynose} and TEMPY \cite{tempy2022}. Each adopts its own techniques and approaches to detect and categorize test smells. To count the test smells, we first summed up all the detections. Then, for each LLM, we apply the calculation of the individual sum of detections by type (see \autoref{eq:3}), allowing for a detailed and segmented analysis of the results.

\begin{equation}\label{eq:3}
    T_{\text{smells}} = \sum_{i=1}^{n} \text{Smell}_{i}
\end{equation}

In total, we found 512 detections of smells in the test suites. After counting all the smells, we calculated the percentage distribution of smells by type (see \autoref{eq:4}). The aim of this equation is to determine the distribution of test smells by type based on the total smells detected in each LLM (see \ref{eq:3}) in relation to the overall number of test smells identified across all LLMs.

\begin{equation}\label{eq:4}
    Distrib_{\text{smells}} = \frac{T_{\text{smells}}}{\sum_{\text{LLMs}} T_{\text{smells}}}
\end{equation}

The test smells detected in this study are categorized in Table \ref{smells-tipo}, which shows the type of test smell detected, the tool that detected it and a brief description. Some test smells were detected by both tools, but they were only counted for the calculation, ruling out the possibility of duplicates.

\begin{table}[H]
\footnotesize
\caption{Test smells detected by tools}
\label{smells-tipo}
\begin{tabular}{lll}
\rowcolor[HTML]{EFEFEF} 
\textbf{Test smell}                                                       & \textbf{Tools}                                            & \textbf{Description}                                                                                                                                                                                                                                                \\
Assertion Roulette                                                        & Pynose                                                    & \begin{tabular}[c]{@{}l@{}}Several asserts without any explanation \\ or message \cite{pynose}.\end{tabular}                                                                                                                                                            \\
\rowcolor[HTML]{EFEFEF} 
\begin{tabular}[c]{@{}l@{}}Lack of Cohesion \\ of Test Cases\end{tabular} & Pynose                                                    & \begin{tabular}[c]{@{}l@{}}Test cases that are grouped together in \\ a test suite, but are not cohesive. Class \\ cohesion is a metric that indicates how \\ well the various parts and responsibilities \\ of a class are joined together \cite{pynose}.\end{tabular} \\
Suboptimal Assertion                                                      & Pynose                                                    & \begin{tabular}[c]{@{}l@{}}Presence of one of the suboptimal \\ asserts \cite{pynose}\end{tabular}                                                                                                                                                                                                                \\
\rowcolor[HTML]{EFEFEF} 
\begin{tabular}[c]{@{}l@{}}Programming \\ Paradigms Blend\end{tabular}    & TEMPY                                                     & \begin{tabular}[c]{@{}l@{}}Mixing paradigms in the same test \\ file \cite{tempy2022}.\end{tabular}                                                                                                                                                                                                                \\
\begin{tabular}[c]{@{}l@{}}Non-Functional \\ Statement\end{tabular}       & TEMPY                                                     & Empty scope in a test method \cite{tempy2022}                                                                                                                                                                                                                         \\
\rowcolor[HTML]{EFEFEF} 
\begin{tabular}[c]{@{}l@{}}Magic Number \\ Test\end{tabular}              & Pynose                                                    & \begin{tabular}[c]{@{}l@{}}Existence of literal numeric values in a \\ test \cite{pynose}.\end{tabular}                                                                                                                                                                 \\
Unknow Test                                                               & TEMPY                                                     & Tests without assertions \cite{tempy2022} .                                                                                                                                                                                                                             \\
\rowcolor[HTML]{EFEFEF} 
\begin{tabular}[c]{@{}l@{}}Conditional \\ Test Logic\end{tabular}         & \begin{tabular}[c]{@{}l@{}}Pynose,  \\ TEMPY\end{tabular} & \begin{tabular}[c]{@{}l@{}}Tests with control statements (if, for, \\ while...) \cite{pynose}.\end{tabular}                                                                                                                                                                 \\
Duplicate Assert                                                          & Pynose                                              & \begin{tabular}[c]{@{}l@{}}Duplicate assertions in the same test \cite{pynose}.\end{tabular}                         
\end{tabular}
\end{table}

\subsubsection{Co-occurrence analysis}
The final stage of the study consisted of analyzing the correlation between the errors detected and the test smells identified by the tools. To do this, only codes that simultaneously presented errors and test smells in the same test suite were considered. The number of test smells detected in each test suite with an error was counted, and a relationship was established between the types of test smells identified and the types of errors reported by the Python interpreter. With this approach, it was possible to assess the co-occurrence of errors and test smells, analyzing both the frequency and distribution of each type within the affected test suites. To count the co-occurrences, we consider the sum (see Equation \ref{eq:6}):

\begin{equation}\label{eq:5}
    k = \sum_{\text{LLMs}} T_{\text{smells}}
\end{equation}

\begin{equation}\label{eq:6}
    Qtd_{\text{co-occurrence}}(S, E) = \sum_{i=1}^{k} \mathbb{1}(S \cap E)
\end{equation}

Where \(k\) is the total number of cases evaluated (total number of smells detected), and \(1(S \cap E)\) is an indicator function that returns 1 if \(S\) (smell) and \(E\) (error) co-occur in the same case, and 0 otherwise. Finally, a total of 512 cases were analyzed, resulting in 265 co-occurrences.

\section{Results}

\begin{table*}[ht]
\footnotesize
\caption{Distribution of errors in test suites generated by LLMs}
\label{tab:errors-distribuicao}
\begin{tabular}{llrrrrrrrrrrrr}
\rowcolor[HTML]{EFEFEF} 
\textbf{LLM}   & \textbf{\begin{tabular}[c]{@{}l@{}}Prompt \\ Context\end{tabular}} & \multicolumn{1}{l}{\cellcolor[HTML]{EFEFEF}\textbf{\begin{tabular}[c]{@{}l@{}}Assertion\\ Error\end{tabular}}} & \multicolumn{1}{l}{\cellcolor[HTML]{EFEFEF}\textbf{\begin{tabular}[c]{@{}l@{}}Indentation\\ Error\end{tabular}}} & \multicolumn{1}{l}{\cellcolor[HTML]{EFEFEF}\textbf{\begin{tabular}[c]{@{}l@{}}Syntax\\ Error\end{tabular}}} & \multicolumn{1}{l}{\cellcolor[HTML]{EFEFEF}\textbf{\begin{tabular}[c]{@{}l@{}}Key\\ Error\end{tabular}}} & \multicolumn{1}{l}{\cellcolor[HTML]{EFEFEF}\textbf{\begin{tabular}[c]{@{}l@{}}Attribute\\ Error\end{tabular}}} & \multicolumn{1}{l}{\cellcolor[HTML]{EFEFEF}\textbf{\begin{tabular}[c]{@{}l@{}}Name\\ Error\end{tabular}}} & \multicolumn{1}{l}{\cellcolor[HTML]{EFEFEF}\textbf{\begin{tabular}[c]{@{}l@{}}Type\\ Error\end{tabular}}} & \multicolumn{1}{l}{\cellcolor[HTML]{EFEFEF}\textbf{\begin{tabular}[c]{@{}l@{}}Overflow\\ Error\end{tabular}}} & \multicolumn{1}{l}{\cellcolor[HTML]{EFEFEF}\textbf{\begin{tabular}[c]{@{}l@{}}Zero\\ Division\\ Error\end{tabular}}} & \multicolumn{1}{l}{\cellcolor[HTML]{EFEFEF}\textbf{\begin{tabular}[c]{@{}l@{}}Index\\ Error\end{tabular}}} & \multicolumn{1}{l}{\cellcolor[HTML]{EFEFEF}\textbf{\begin{tabular}[c]{@{}l@{}}Value\\ Error\end{tabular}}} & \multicolumn{1}{l}{\cellcolor[HTML]{EFEFEF}\textbf{TOTAL}} \\
\rowcolor[HTML]{FFFFFF} 
GPT-4o & C2C                                                                & 0,112                                                                                                          & 0,000                                                                                                            & 0,286                                                                                                       & 0,143                                                                                                    & 0,000                                                                                                          & 0,000                                                                                                     & 0,000                                                                                                     & 0,333                                                                                                         & 0,000                                                                                                                & 0,000                                                                                                      & 1,000                                                                                                      & 0,106                                                      \\
\rowcolor[HTML]{EFEFEF} 
GPT-4o & T2C                                                                & 0,082                                                                                                          & 0,000                                                                                                            & 0,000                                                                                                       & 0,000                                                                                                    & 0,000                                                                                                          & 0,000                                                                                                     & 0,000                                                                                                     & 0,333                                                                                                         & 0,500                                                                                                                & 0,000                                                                                                      & 0,000                                                                                                      & 0,066                                                      \\
\rowcolor[HTML]{FFFFFF} 
Amazon Q       & C2C                                                                & 0,194                                                                                                          & 0,235                                                                                                            & 0,143                                                                                                       & 0,143                                                                                                    & 0,000                                                                                                          & 0,600                                                                                                     & 0,000                                                                                                     & 0,333                                                                                                         & 0,000                                                                                                                & 0,000                                                                                                      & 0,000                                                                                                      & 0,192                                                      \\
\rowcolor[HTML]{EFEFEF} 
Amazon Q       & T2C                                                                & 0,184                                                                                                          & 0,765                                                                                                            & 0,286                                                                                                       & 0,286                                                                                                    & 0,667                                                                                                          & 0,400                                                                                                     & 0,500                                                                                                     & 0,000                                                                                                         & 0,000                                                                                                                & 0,000                                                                                                      & 0,000                                                                                                      & 0,285                                                      \\
\rowcolor[HTML]{FFFFFF} 
LLama          & C2C                                                                & 0,276                                                                                                          & 0,000                                                                                                            & 0,000                                                                                                       & 0,143                                                                                                    & 0,167                                                                                                          & 0,000                                                                                                     & 0,500                                                                                                     & 0,000                                                                                                         & 0,500                                                                                                                & 1,000                                                                                                      & 0,000                                                                                                      & 0,219                                                      \\
\rowcolor[HTML]{EFEFEF} 
LLama          & T2C                                                                & 0,153                                                                                                          & 0,000                                                                                                            & 0,286                                                                                                       & 0,286                                                                                                    & 0,167                                                                                                          & 0,000                                                                                                     & 0,000                                                                                                     & 0,000                                                                                                         & 0,000                                                                                                                & 0,000                                                                                                      & 0,000                                                                                                      & 0,132                                                      \\
\rowcolor[HTML]{FFFFFF} 
\multicolumn{2}{r}{\cellcolor[HTML]{FFFFFF}\textbf{Distribution \%}}                & \textbf{0,649}                                                                                                 & \textbf{0,113}                                                                                                   & \textbf{0,046}                                                                                              & \textbf{0,046}                                                                                           & \textbf{0,040}                                                                                                 & \textbf{0,033}                                                                                            & \textbf{0,026}                                                                                            & \textbf{0,020}                                                                                                & \textbf{0,013}                                                                                                       & \textbf{0,007}                                                                                             & \textbf{0,007}                                                                                             & \textbf{1,000}                                            
\end{tabular}
\end{table*}

\subsection{\textbf{RQ$_1$: Types and Frequency of Errors}}

To analyze the errors, all the test suites generated were run together with their respective production codes, previously extracted from HumanEval. Tests that ran successfully, were classified as valid. Those that failed or resulted in exceptions raised by the Python interpreter were recorded as error cases. In total, 151 errors were identified in the test suites analyzed. It is important to note that some suites had multiple errors, and each occurrence was counted separately for analysis purposes, reflecting the diversity and frequency of the faults detected.

Table \ref{tab:errors-distribuicao} shows the distribution of the types of errors detected in the test suites for the C2C and T2C contexts, analyzed in each of the three LLMs. Overall, assertion errors (\textit{AssertionError}) were the most prevalent, accounting for 64.9\% of the total. This result shows that many tests were generated with incorrect expectations regarding the behavior of the production code. The second most frequent type of error was \textit{IndentationError}, which accounted for 11.3\% of detections, indicating that several parts of the generated tests did not follow the indentation standards required by Python, resulting in execution failures. \textit{SyntaxError} and \textit{KeyError} errors had the same detection rate (4.6\%), suggesting that some suites had syntax errors or cases of “hallucinations”, where LLMs tried to access non-existent variables or functions. Other errors were also detected to a lesser extent: \textit{AttributeError} (4\%), \textit{NameError} (3.3\%), \textit{TypeError} (2.6\%), \textit{OverflowError} (2\%), \textit{ZeroDivisionError} (1.3\%), and, less frequently, \textit{IndexError} and \textit{ValueError} (both 0.7\%). 

When analyzing each LLM studied individually (Table \ref{tab:errors-distribuicao}), it can be seen that \texttt{GPT-4o} was the model with the lowest incidence of errors, totaling 10.6\% and 6.6\% of detections in the C2C and T2C contexts, respectively. In addition, \texttt{GPT-4o} had the lowest diversity of errors by type, with most occurrences concentrated in \textit{AssertionError} and \textit{SyntaxError}. On the other hand, \texttt{Amazon Q} Developer recorded a high error rate in the test suites, corresponding to 19.2\% of detections in the C2C context and 28.5\% in T2C. This model was also the only one responsible for \textit{IndentationError} cases, due to many generated tests not being completed correctly. In several cases, test functions were declared with inadequate indentation, compromising their execution. \texttt{LLaMA}, on the other hand, had an intermediate performance, with 21.9\% of errors in the C2C context and 13.2\% in T2C. These results place \texttt{LLaMA} between the two extremes, showing a moderate distribution of errors compared to the other models.

\textit{\textbf{Answer to RQ$_1$:}} We identified a total of 151 occurrences of errors in the test suites generated by the LLMs. Among these errors, the most frequent were \textit{AssertionError} (64.9\%), \textit{IndentationError} (11.3\%) and \textit{SyntaxError} (4.6\%). When analyzing the results by LLM, \texttt{GPT-4o} had the fewest errors, while \texttt{Amazon Q} Developer had the most. Notably, some types of errors, such as \textit{IndentationError} and \textit{NameError}, were exclusive to the tests generated by \texttt{Amazon Q} Developer. \texttt{LLama}, on the other hand, although it had an intermediate occurrence of errors overall, was responsible for the highest number of \textit{AssertionError} errors. This suggests that certain models may be more reliable than others in generating valid tests, but also that specific errors, such as assertion errors, may be more prevalent in some cases, affecting the accuracy and reliability of the tests generated, since errors of this type were detected in the tests of all three LLMs.

\subsection{\textbf{RQ$_2$: Types and Distribution of Test Smells}}

\begin{table*}[ht]
\footnotesize
\caption{Distribution of test smells in test suites generated by LLMs}
\label{tab:smells-detection}
\begin{tabular}{llrrrrrrrrrr}
\rowcolor[HTML]{EFEFEF} 
\textbf{LLM}                                              & \textbf{\begin{tabular}[c]{@{}l@{}}Prompt \\ Context\end{tabular}} & \multicolumn{1}{l}{\cellcolor[HTML]{EFEFEF}\textbf{\begin{tabular}[c]{@{}l@{}}Lack of \\ Cohesion \\ of Test Cases\end{tabular}}} & \multicolumn{1}{l}{\cellcolor[HTML]{EFEFEF}\textbf{\begin{tabular}[c]{@{}l@{}}Assertion \\ Roulette\end{tabular}}} & \multicolumn{1}{l}{\cellcolor[HTML]{EFEFEF}\textbf{\begin{tabular}[c]{@{}l@{}}Programming \\ Paradigms \\ Blend\end{tabular}}} & \multicolumn{1}{l}{\cellcolor[HTML]{EFEFEF}\textbf{\begin{tabular}[c]{@{}l@{}}Suboptimal \\ Assertion\end{tabular}}} & \multicolumn{1}{l}{\cellcolor[HTML]{EFEFEF}\textbf{\begin{tabular}[c]{@{}l@{}}Non-\\ Functional \\ Statement\end{tabular}}} & \multicolumn{1}{l}{\cellcolor[HTML]{EFEFEF}\textbf{\begin{tabular}[c]{@{}l@{}}Conditional \\ Test Logic\end{tabular}}} & \multicolumn{1}{l}{\cellcolor[HTML]{EFEFEF}\textbf{\begin{tabular}[c]{@{}l@{}}Unknown \\ Test\end{tabular}}} & \multicolumn{1}{l}{\cellcolor[HTML]{EFEFEF}\textbf{\begin{tabular}[c]{@{}l@{}}Duplicated \\ Assert\end{tabular}}} & \multicolumn{1}{l}{\cellcolor[HTML]{EFEFEF}\textbf{\begin{tabular}[c]{@{}l@{}}Exception \\ Handling\end{tabular}}} & \multicolumn{1}{l}{\cellcolor[HTML]{EFEFEF}\textbf{TOTAL}} \\
\rowcolor[HTML]{FFFFFF} 
GPT-4o & C2C                                                                & 0,171                                                                                                                             & 0,185                                                                                                              & 0,012                                                                                                                          & 0,056                                                                                                                & 0,000                                                                                                                       & 0,000                                                                                                                  & 0,000                                                                                                        & 0,000                                                                                                             & 0,000                                                                                                              & 0,121                                                      \\
\rowcolor[HTML]{EFEFEF} 
GPT-4o & T2C                                                                & 0,180                                                                                                                             & 0,137                                                                                                              & 0,432                                                                                                                          & 0,917                                                                                                                & 0,000                                                                                                                       & 0,000                                                                                                                  & 0,000                                                                                                        & 0,000                                                                                                             & 0,000                                                                                                              & 0,240                                                      \\
\rowcolor[HTML]{FFFFFF} 
\begin{tabular}[c]{@{}l@{}}Amazon \\ Q\end{tabular}                                                  & C2C                                                                & 0,104                                                                                                                             & 0,097                                                                                                              & 0,099                                                                                                                          & 0,000                                                                                                                & 0,000                                                                                                                       & 0,000                                                                                                                  & 0,316                                                                                                        & 0,143                                                                                                             & 1,000                                                                                                              & 0,098                                                      \\
\rowcolor[HTML]{EFEFEF} 
\begin{tabular}[c]{@{}l@{}}Amazon \\ Q\end{tabular}                                                  & T2C                                                                & 0,180                                                                                                                             & 0,315                                                                                                              & 0,444                                                                                                                          & 0,028                                                                                                                & 1,000                                                                                                                       & 0,143                                                                                                                  & 0,684                                                                                                        & 0,857                                                                                                             & 0,000                                                                                                              & 0,301                                                      \\
\rowcolor[HTML]{FFFFFF} 
LLama                                                     & C2C                                                                & 0,185                                                                                                                             & 0,202                                                                                                              & 0,012                                                                                                                          & 0,000                                                                                                                & 0,000                                                                                                                       & 0,357                                                                                                                  & 0,000                                                                                                        & 0,000                                                                                                             & 0,000                                                                                                              & 0,137                                                      \\
\rowcolor[HTML]{EFEFEF} 
LLama                                                     & T2C                                                                & 0,180                                                                                                                             & 0,065                                                                                                              & 0,000                                                                                                                          & 0,000                                                                                                                & 0,000                                                                                                                       & 0,500                                                                                                                  & 0,000                                                                                                        & 0,000                                                                                                             & 0,000                                                                                                              & 0,104                                                      \\
\rowcolor[HTML]{FFFFFF} 
\multicolumn{2}{r}{\cellcolor[HTML]{FFFFFF}\textbf{Distribution \%}}                                                           & \textbf{0,412}                                                                                                                    & \textbf{0,242}                                                                                                     & \textbf{0,158}                                                                                                                 & \textbf{0,070}                                                                                                       & \textbf{0,037}                                                                                                              & \textbf{0,027}                                                                                                         & \textbf{0,037}                                                                                               & \textbf{0,014}                                                                                                    & \textbf{0,002}                                                                                                     & \textbf{1,000}                                            
\end{tabular}
\end{table*}

Each test suite was analyzed using specialized tools for detecting test smells. All detections made by the two tools were counted, considering each smell identified individually, even when multiple smells were found in the same test suite. By consolidating the results from both tools, 512 detections of test smells were recorded, which served as the basis for the analysis in this study.

The results of the distribution of test smells are shown in Table \ref{tab:smells-detection}. The analysis revealed that the most frequently identified smell was \textit{Lack of Cohesion of Test Cases}, responsible for 41.2\% of all detections. This indicates that a significant number of test suites contained cases that lacked cohesion between them, compromising the structural quality of the tests. Next, the \textit{Assertion Roulette} smell was detected in 24.2\% of cases, which suggests that many tests had multiple assertions without clear documentation or explanations, making them difficult to understand and maintain. \textit{Programming Paradigms Blend} was the third most frequent smell, with 15.8\% of detections, indicating that some tests used fields initialized outside the test class (in global scope), which is considered an inappropriate practice. Another relevant smell was the \textit{Suboptimal Assertion}, documented by \citet{pynose} in the Pynose tool, which obtained recurrent detections. This smell reflects assertions that were not specific, robust or relevant enough to validate the expected behavior of the production code. In addition, Table \ref{tab:smells-detection} highlights other less prevalent test smells: \textit{Non-Functional Statement} and \textit{Unknown Test} appeared in 3.7\% of detections, \textit{Conditional Test Logic} in 2.7\%, \textit{Duplicated Assert} in 1.4\%, and \textit{Exception Handling} was identified only once. These results reinforce the diversity and complexity of the design problems found in the test suites generated by LLMs.

In the analysis by LLM, Amazon Q Developer showed the greatest recurrence and diversity of test smells, standing out as the model with the greatest quantity and distribution of detections. Notably, some smells were exclusive to the code generated by this model, such as \textit{Unknown Test, Non-Functional Statement, Duplicated Assert} and \textit{Exception Handling}, which shows specific problems in the quality of the tests produced. On the other hand, LLama showed the best performance in terms of structural quality, with the lowest number and variety of smells detected. Only three types were identified in its test suites: \textit{Lack of Cohesion of Test Cases, Assertion Roulette} and \textit{Conditional Test Logic}, demonstrating greater consistency in generating tests with fewer quality violations. GPT-4o fell between these two extremes, with a moderate volume of detections and a low diversity of test smells. Only four types were identified, similar to LLama. However, there was a significant concentration of the \textit{Suboptimal Assertion smell}, which had a high frequency of detections in one of the contexts analyzed. This pattern suggests that although GPT-4o produces less variety of smells, it may be prone to specific issues related to the quality of the assertions generated.

\textit{\textbf{Answer to RQ$_2$:}} We identified a total of 512 detections of test smells in the test suites generated by the LLMs. \textit{Lack of Cohesion of Test Cases} was the most frequent smell (41.2\%), followed by Assertion Roulette (24.2\%) and \textit{Programming Paradigms Blend} (15.8\%). Other less prevalent smells included \textit{Non-Functional Statement, Unknown Test, Conditional Test Logic, Duplicated Assert,} and \textit{Exception Handling}. Among the LLMs, Amazon Q Developer exhibited the highest occurrence and diversity of test smells, with several issues being unique to its generated code, such as \textit{Unknown Test} and \textit{Non-Functional Statement}. In contrast, LLama had the best structural quality, with only three types of smells detected, suggesting more cohesive and well-structured tests. GPT-4o showed moderate performance, with four types of smells identified but a notable concentration in \textit{Suboptimal Assertion}, indicating frequent issues with assertion specificity and robustness.

\subsection{\textbf{RQ$_3$: Prompt Influence}}

Based on the analysis of Tables \ref{tab:errors-distribuicao} and \ref{tab:smells-detection}, it was possible to observe the impact of the different prompt types on the occurrence of errors and test smells in the test suites generated by each LLM. Based on this data, we conducted a detailed analysis, separating the total number of detections for each prompt context, allowing for a more in-depth understanding of the influence of context on the quality of the generations.

\subsubsection{Analyzing context Code2Code}

\begin{figure}[ht]
    \centering 
    \includegraphics[width=1\linewidth]{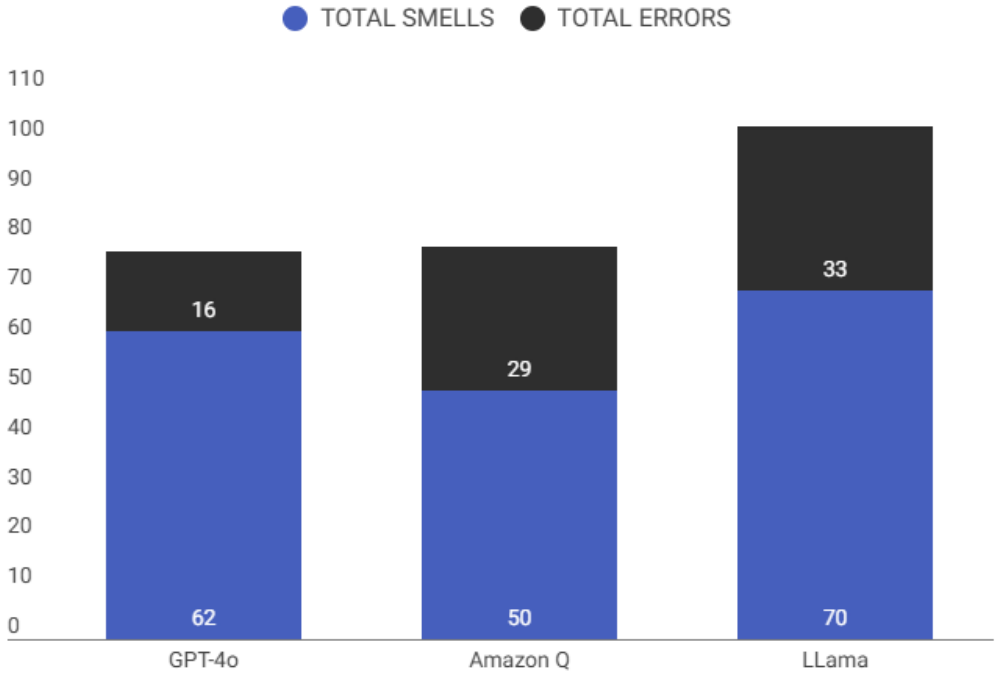} 
    \caption{Quantity of errors and test smells in the C2C context}
    \Description{}
    \label{fig:graph1}
\end{figure}

When analyzing the data from the test suites generated in the C2C context (Figure \ref{fig:graph1}), we observed a moderate variation in the detection of test smells. The Amazon Q model had 50 detections, while LLaMA led with 70, and GPT-4o was in an intermediate position with 62 detections. As for the occurrence of errors, GPT-4o stood out for having the lowest number in this context, while the LLaMA (33) and Amazon Q (29) models had a similar incidence. These results indicate that, despite having generated the tests based on the production code provided in the prompt, the Amazon Q and LLaMA models showed a high occurrence of errors in the tests generated, which shows weaknesses in the quality of their generations in this scenario.

\subsubsection{Analyzing context Text2Code}

\begin{figure}[ht]
    \scriptsize
    \centering 
    \includegraphics[width=1\linewidth]{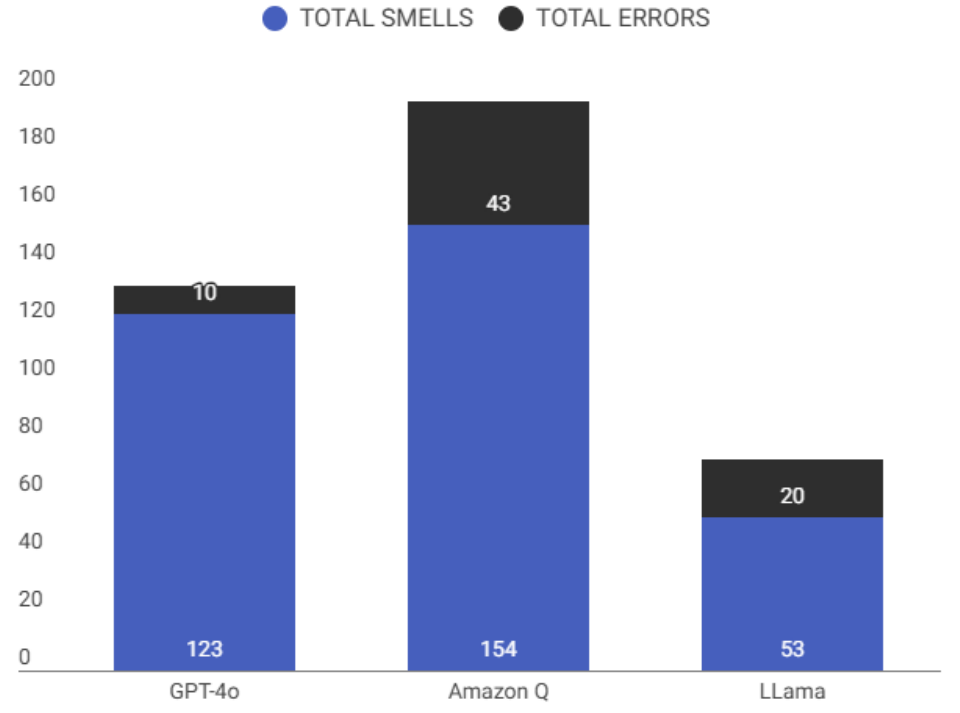} 
    \caption{Quantity of errors and test smells in the T2C context}
    \Description{}
    \label{fig:graph2}
\end{figure}

In the T2C context in Figure \ref{fig:graph2}, the incidence of test smells in GPT-4o doubled compared to the C2C context, reaching 123 detections, although the occurrence of errors was lower. Amazon Q had an even sharper increase, with the number of smells tripling (154 detections) and the incidence of errors rising to 43. On the other hand, LLaMA showed a significant improvement compared to the C2C context, with a reduction in both the number of test smells and the number of errors. These results indicate that LLaMA has shown greater ability to generate quality tests from textual descriptions, outperforming Amazon Q. In addition, Amazon Q faced critical difficulties in interpreting text to generate code, with some of the problems related to errors in the code produced. In the case of GPT-4o, the main challenge in the T2C context was the high incidence of test smells in the structure of the codes generated. However, the number of errors remained low, indicating a functional but poorly optimized structure.

\textit{\textbf{Answer to RQ$_3$:}} The results of this analysis show the individuality of each LLM in relation to the two prompt contexts analyzed. It was not possible to identify a consistent pattern between the three models in terms of the quality of the test suites generated, since each LLM showed different behaviors. 

GPT-4o stood out as the most stable model between the two contexts, with a low incidence of errors in both prompts. However, in the T2C context, there was a significant increase in the number of test smells, which suggests that the model structures the tests better by having direct access to the production code in the prompt. Amazon Q faced major difficulties in the T2C context, with a high incidence of errors and test smells. This indicates that the model is more reliable for generating tests when it can visualize the code directly, as in the C2C context. On the other hand, LLaMA performed better when generating tests in the T2C context, with a considerable reduction in errors and test smells compared to the C2C context. This behavior indicates that the model is better able to interpret textual descriptions and generate quality tests from them.

These observations reinforce the importance of understanding the particularities of each LLM to optimize the use of the different prompt contexts, depending on the specific needs of quality and reliability in test generation.

\subsection{\textbf{RQ$_4$: Correlation between Errors and Test Smells}}

\begin{figure*}[htb]
    \centering 
    \includegraphics[width=1.0\textwidth]{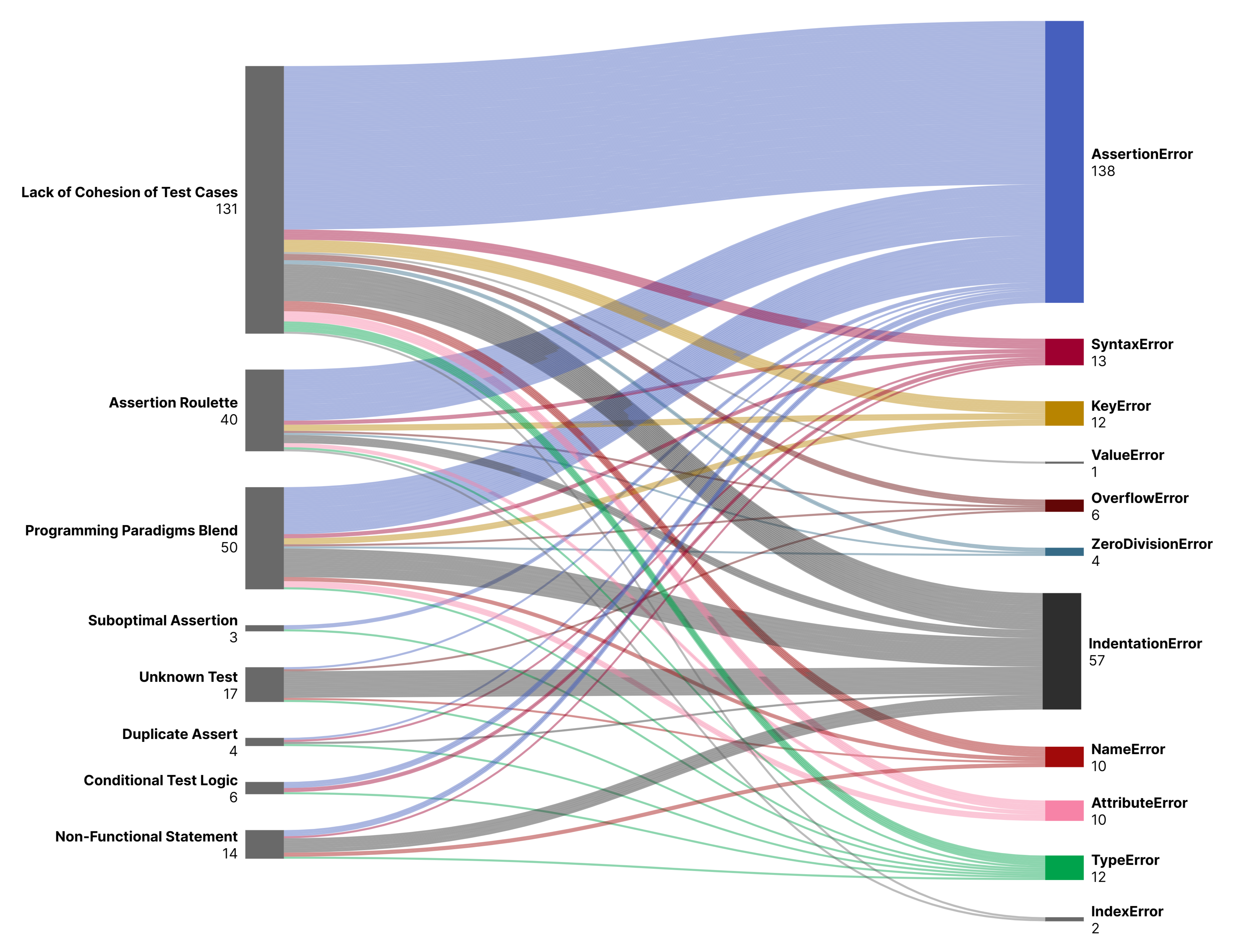} 
    \caption{Cooccurrences of Errors and Test Smells by type}
    \Description{}
    \label{fig:coocorrencia}
\end{figure*}

To correlate the occurrence of errors with test smells, we accounted for all test smell detections and identified those directly associated with errors. This analysis revealed that out of 512 test smell detections, 265 were found in test suites that exhibited some type of error. Figure \ref{fig:coocorrencia} visually demonstrates the relationships between the detected test smells and their associated errors. These results represent the aggregated detections across all three LLMs analyzed in this study.

\textit{Lack of Cohesion of Test Cases} was the most frequent test smell, with 131 occurrences, and is strongly associated with \textit{AssertionError}, which accounted for 138 occurrences. This relationship indicates that test suites with low cohesion often fail to validate the expected behavior of production code. In addition, this test smell also had connections with errors such as \textit{IndentationError} and \textit{SyntaxError}, showing structural and formatting problems in tests with non-cohesive cases. \textit{Assertion Roulette}, identified 40 times, is mostly linked to \textit{AssertionError}, highlighting that tests with multiple assertions without clear explanations considerably increased the likelihood of incorrect expectations. On the other hand, the \textit{Programming Paradigms Blend}, with 50 detections, showed a wide dispersion of connections, including errors such as \textit{IndentationError} and \textit{TypeError}. This suggests that mixing paradigms in tests can damage their structure and logical execution.

\textit{Unknown Test}, with 17 occurrences, stood out for its strong association with \textit{IndentationError}. This connection indicates that many tests classified as unknown had serious formatting problems, related to the fact that Amazon Q Developer did not structure the code properly (as discussed in RQ$_3$). Other test smells, such as \textit{Non-Functional Statement} (14 occurrences), have multiple connections with errors such as \textit{KeyError} and \textit{NameError}. These results indicate that poorly defined or non-functional tests often use non-existent or invalid elements, causing critical failures. In addition, \textit{Duplicate Assert} and \textit{Conditional Test Logic}, although less frequent, are associated with \textit{AssertionError}, highlighting that problems with conditional logic or duplicate assertions directly affect the effectiveness of tests.

The less recurrent errors, such as \textit{OverflowError}, \textit{ZeroDivisionError}, \textit{IndexError} and \textit{ValueError}, appear less frequently, but maintain connections with various test smells. This suggests that specific failures may be caused by a combination of structural and logical problems. Finally, it is important to note that all the test smells detected are directly or indirectly linked to \textit{AssertionError}, reinforcing that incorrect validation is a critical problem in the test suites generated by LLMs.

\textit{\textbf{Answer to RQ$_4$:}} The analysis revealed that 265 of the 512 detections of test smells occurred in tests with errors, showing a strong relationship between structural problems and test failures generated by LLMs.  \textit{Lack of Cohesion of Test Cases} was the most frequent test smell. It showed a strong association with \textit{AssertionError}, indicating that tests with low cohesion often fail to validate the code. In addition, errors such as \textit{IndentationError} and \textit{SyntaxError} were common in non-cohesive tests, indicating structural flaws.  \textit{Assertion Roulette} also increased the incidence of \textit{AssertionError}, showing that multiple assertions without context lead to incorrect expectations. 

The mixture of paradigms in the tests generated a variety of errors, affecting both the structure and logic of the tests. Poorly defined tests or problematic conditional logic also contributed to critical failures. Overall, the predominance of  \textit{AssertionError} in almost all test smells reinforces that inadequate validation is a central problem in tests generated by LLMs, compromising their reliability.

\section{Discussion and Implications}

The results of this study have direct implications for the software testing industry, especially for teams adopting agile methodologies and practices such as Test-Driven Development (TDD). Analyzing the quality of tests generated by LLMs in Code2Code (C2C) and Text2Code (T2C) contexts reveals specific challenges for different developer profiles and test-writing approaches.

In the scenario of developers writing tests directly from code (C2C), as in traditional TDD approaches, automatic test generation can be used to speed up the development cycle. However, our results indicate that LLMs still have limitations in producing cohesive and well-structured tests, with a high incidence of test smells such as  \textit{Lack of Cohesion of Test Cases} and  \textit{Assertion Roulette}.  Thus, manual review is still needed to ensure the quality of the tests generated, as errors such as  \textit{AssertionError} and  \textit{IndentationError} were recurrent, requiring refinement before adoption in a real environment.

On the other hand, developers who write tests based on textual descriptions (T2C), as in teams that follow Behavior-Driven Development (BDD) or document requirements before implementation, may face different challenges.  Our findings indicate that LLMs tend to produce tests more prone to structural errors and have a high incidence of test smells, especially  \textit{Suboptimal Assertions}. For teams working with agile methodologies and prioritizing natural language documentation, the use of LLMs can speed up test creation, but requires strategies to mitigate the low specificity of the assertions generated.

In addition, we observed differences in the performance of the LLMs in the different scenarios analyzed. GPT-4o, for example, had the lowest error rate and the greatest consistency in test generation, making it a viable option for teams seeking greater reliability in test automation. However, it had a high incidence of Suboptimal Assertions, which could impact teams using TDD or BDD, as it would require adjustments to the verifications of each test generated. Amazon Q Developer, on the other hand, showed a high recurrence of structural errors, such as  \textit{IndentationError} and  \textit{NameError}, as well as several unique test smells. This suggests that its use may be more limited for scenarios that require well-structured tests. LLama had an intermediate performance, with a moderate number of errors, but stood out negatively for having the highest number of  \textit{AssertionError}. On the other hand, this model performed well when generating tests from text (T2C), which could be a plausible choice for developers working with documentation or BDD.

The adoption of LLMs for automatic test generation needs to be accompanied by good practices, such as manual review, more refined prompt engineering, and integration with static analysis tools. Companies using agile methodologies can benefit from these technologies to speed up test writing, but they need to consider the risks associated with the reliability of the tests generated, ensuring that automation does contribute to the continuous improvement of software quality.

\section{Threats to Validity}

\begin{itemize}
    \item \textbf{Conclusion Validity} Although the tests were generated from the HumanEval dataset, the diversity of scenarios may not be wide enough to capture all the possible faults and test smells that LLMs can produce. Detecting errors and test smells alone may not be enough to accurately determine the quality of the tests generated. In addition, the analysis of these problems depends on automated tools and the language used, which have limitations and may not identify all the inconsistencies present in the tests.
    
    \item \textbf{Internal Validity}
    Variability in code generation by LLMs can influence the reproducibility of tests, especially if the models are updated in the future. In addition, possible differences in the parameters used by the LLMs may have impacted the quality of the tests generated, making it difficult to guarantee that the results are completely attributable to the types of prompts analyzed.
    
    \item \textbf{Construct Validity}
    Specific tools were used to detect test smells (Pynose and TEMPY), but they may not cover all smells in automated testing. In addition, the categorization of errors was based on messages from the Python interpreter, which can lead to different interpretations of the criticality of the problems encountered.
    
    \item \textbf{External Validity}
    The results may not be generalizable to other software domains, as the study was conducted exclusively with Python code and tests. In addition, only three LLMs were analyzed (GPT-4o, Amazon Q Developer and LLama 3.3), which may limit the applicability of the findings to other code generation models. The production codes present in HumanEval cover specific programming problems that are often used to train LLMs. As such, our results may not be generalizable to other code domains outside this scope, limiting the applicability of the findings to more complex or diverse contexts.
\end{itemize}

\section{Conclusion and Future Work}

In this paper, we conducted a comparative study of the quality of tests generated by three LLMs (GPT-4o, Amazon Q Developer and LLama 3.3). We used the HumanEval dataset to generate tests from prompts in the Text2Code and Code2Code contexts. After generation, the test suites were run alongside production code to check for errors. We then used test smells detection tools to assess the structural quality of the generated tests. In addition, we correlated the presence of errors and test smells in the same test suite.

Our results highlight the influence of the type of prompt and the LLM model on the quality of the test suites generated. The analysis revealed that the most frequent errors were  \textit{AssertionError, IndentationError} and  \textit{SyntaxError}, with significant variations between the models evaluated. GPT-4o had the lowest error rate, while Amazon Q Developer was the most likely to generate tests with flaws, including  \textit{IndentationError} and  \textit{NameError}, which were unique to this model. In addition, we identified that the most prevalent test smells were  \textit{Lack of Cohesion of Test Cases} and  \textit{Assertion Roulette}, indicating that many suites generated by LLMs suffer from low cohesion and multiple assertions without adequate contextualization. The correlation between errors and test smells showed that structural problems have a direct impact on the reliability of automated tests.

Despite advances in automatic test generation, our findings suggest that LLMs still face challenges in producing high-quality suites, mainly due to the recurring presence of errors and violations of good testing practices. In view of this, in future studies we intend to explore different prompt engineering approaches, broaden the diversity of datasets evaluated and consider additional metrics for a more comprehensive analysis of the quality of tests generated by LLMs.

\section*{Artifact Availability}
We provide our data and artifacts under open licenses at \textit{\url{https://github.com/ovictorpa/llms-research}} and \textit{\url{https://zenodo.org/records/14773297}}

\begin{acks}
This study was financed in part by the Coordenação de Aperfeiçoamento de Pessoal de Nível Superior - Brasil (CAPES) - Finance Code 001; FAPESB grant PIE0002/2022; and CNPq grants 315840/2023-4 and 403361/2023-0.
\end{acks}


\begin{thebibliography}{45}


\ifx \showCODEN    \undefined \def \showCODEN     #1{\unskip}     \fi
\ifx \showDOI      \undefined \def \showDOI       #1{#1}\fi
\ifx \showISBNx    \undefined \def \showISBNx     #1{\unskip}     \fi
\ifx \showISBNxiii \undefined \def \showISBNxiii  #1{\unskip}     \fi
\ifx \showISSN     \undefined \def \showISSN      #1{\unskip}     \fi
\ifx \showLCCN     \undefined \def \showLCCN      #1{\unskip}     \fi
\ifx \shownote     \undefined \def \shownote      #1{#1}          \fi
\ifx \showarticletitle \undefined \def \showarticletitle #1{#1}   \fi
\ifx \showURL      \undefined \def \showURL       {\relax}        \fi
\providecommand\bibfield[2]{#2}
\providecommand\bibinfo[2]{#2}
\providecommand\natexlab[1]{#1}
\providecommand\showeprint[2][]{arXiv:#2}

\bibitem[Agarwal et~al\mbox{.}(2024)]%
        {agarwal2024structuredcoderepresentationsenable}
\bibfield{author}{\bibinfo{person}{Mayank Agarwal}, \bibinfo{person}{Yikang Shen}, \bibinfo{person}{Bailin Wang}, \bibinfo{person}{Yoon Kim}, {and} \bibinfo{person}{Jie Chen}.} \bibinfo{year}{2024}\natexlab{}.
\newblock \bibinfo{title}{Structured Code Representations Enable Data-Efficient Adaptation of Code Language Models}.
\newblock
\newblock
\showeprint[arxiv]{2401.10716}~[cs.CL]
\urldef\tempurl%
\url{https://arxiv.org/abs/2401.10716}
\showURL{%
\tempurl}


\bibitem[Alshahwan et~al\mbox{.}(2024)]%
        {nadia2024}
\bibfield{author}{\bibinfo{person}{Nadia Alshahwan}, \bibinfo{person}{Jubin Chheda}, \bibinfo{person}{Anastasia Finogenova}, \bibinfo{person}{Beliz Gokkaya}, \bibinfo{person}{Mark Harman}, \bibinfo{person}{Inna Harper}, \bibinfo{person}{Alexandru Marginean}, \bibinfo{person}{Shubho Sengupta}, {and} \bibinfo{person}{Eddy Wang}.} \bibinfo{year}{2024}\natexlab{}.
\newblock \showarticletitle{Automated Unit Test Improvement using Large Language Models at Meta}. In \bibinfo{booktitle}{\emph{Companion Proceedings of the 32nd ACM International Conference on the Foundations of Software Engineering}} (Porto de Galinhas, Brazil) \emph{(\bibinfo{series}{FSE 2024})}. \bibinfo{publisher}{Association for Computing Machinery}, \bibinfo{address}{New York, NY, USA}, \bibinfo{pages}{185–196}.
\newblock
\showISBNx{9798400706585}
\urldef\tempurl%
\url{https://doi.org/10.1145/3663529.3663839}
\showDOI{\tempurl}


\bibitem[Alves et~al\mbox{.}(2024)]%
        {alves2024sbes}
\bibfield{author}{\bibinfo{person}{Victor Alves}, \bibinfo{person}{Cristiano Santos}, \bibinfo{person}{Carla Bezerra}, {and} \bibinfo{person}{Ivan Machado}.} \bibinfo{year}{2024}\natexlab{}.
\newblock \showarticletitle{Detecting Test Smells in Python Test Code Generated by LLM: An Empirical Study with GitHub Copilot}. In \bibinfo{booktitle}{\emph{Anais do XXXVIII Simpósio Brasileiro de Engenharia de Software}} (Curitiba/PR). \bibinfo{publisher}{SBC}, \bibinfo{address}{Porto Alegre, RS, Brasil}, \bibinfo{pages}{581--587}.
\newblock
\showISSN{2833-0633}
\urldef\tempurl%
\url{https://doi.org/10.5753/sbes.2024.3561}
\showDOI{\tempurl}


\bibitem[Aranda et~al\mbox{.}(2024)]%
        {manoel2024}
\bibfield{author}{\bibinfo{person}{Manoel Aranda}, \bibinfo{person}{Naelson Oliveira}, \bibinfo{person}{Elvys Soares}, \bibinfo{person}{M\'{a}rcio Ribeiro}, \bibinfo{person}{Davi Rom\~{a}o}, \bibinfo{person}{Ullyanne Patriota}, \bibinfo{person}{Rohit Gheyi}, \bibinfo{person}{Emerson Souza}, {and} \bibinfo{person}{Ivan Machado}.} \bibinfo{year}{2024}\natexlab{}.
\newblock \showarticletitle{A Catalog of Transformations to Remove Smells From Natural Language Tests}. In \bibinfo{booktitle}{\emph{Proceedings of the 28th International Conference on Evaluation and Assessment in Software Engineering}} (Salerno, Italy) \emph{(\bibinfo{series}{EASE '24})}. \bibinfo{publisher}{Association for Computing Machinery}, \bibinfo{address}{New York, NY, USA}, \bibinfo{pages}{7–16}.
\newblock
\showISBNx{9798400717017}
\urldef\tempurl%
\url{https://doi.org/10.1145/3661167.3661225}
\showDOI{\tempurl}


\bibitem[Bodea(2022)]%
        {pytest-smell2022}
\bibfield{author}{\bibinfo{person}{Alexandru Bodea}.} \bibinfo{year}{2022}\natexlab{}.
\newblock \showarticletitle{Pytest-Smell: a smell detection tool for Python unit tests}. In \bibinfo{booktitle}{\emph{Proceedings of the 31st ACM SIGSOFT International Symposium on Software Testing and Analysis}} (Virtual, South Korea) \emph{(\bibinfo{series}{ISSTA 2022})}. \bibinfo{publisher}{Association for Computing Machinery}, \bibinfo{address}{New York, NY, USA}, \bibinfo{pages}{793–796}.
\newblock
\showISBNx{9781450393799}
\urldef\tempurl%
\url{https://doi.org/10.1145/3533767.3543290}
\showDOI{\tempurl}


\bibitem[Cain(2024)]%
        {Cain2024}
\bibfield{author}{\bibinfo{person}{W. Cain}.} \bibinfo{year}{2024}\natexlab{}.
\newblock In \bibinfo{booktitle}{\emph{Prompting Change: Exploring Prompt Engineering in Large Language Model AI and Its Potential to Transform Education}}, Vol.~\bibinfo{volume}{68}. \bibinfo{publisher}{TechTrends}, \bibinfo{pages}{47--57}.
\newblock
\urldef\tempurl%
\url{https://doi.org/10.1007/s11528-023-00896-0}
\showDOI{\tempurl}


\bibitem[Chen et~al\mbox{.}(2021)]%
        {codex2021}
\bibfield{author}{\bibinfo{person}{Mark Chen}, \bibinfo{person}{Jerry Tworek}, \bibinfo{person}{Heewoo Jun}, \bibinfo{person}{Qiming Yuan}, \bibinfo{person}{Henrique~Pond{\'{e}} de Oliveira~Pinto}, \bibinfo{person}{Jared Kaplan}, \bibinfo{person}{Harrison Edwards}, \bibinfo{person}{Yuri Burda}, \bibinfo{person}{Nicholas Joseph}, \bibinfo{person}{Greg Brockman}, \bibinfo{person}{Alex Ray}, \bibinfo{person}{Raul Puri}, \bibinfo{person}{Gretchen Krueger}, \bibinfo{person}{Michael Petrov}, \bibinfo{person}{Heidy Khlaaf}, \bibinfo{person}{Girish Sastry}, \bibinfo{person}{Pamela Mishkin}, \bibinfo{person}{Brooke Chan}, \bibinfo{person}{Scott Gray}, \bibinfo{person}{Nick Ryder}, \bibinfo{person}{Mikhail Pavlov}, \bibinfo{person}{Alethea Power}, \bibinfo{person}{Lukasz Kaiser}, \bibinfo{person}{Mohammad Bavarian}, \bibinfo{person}{Clemens Winter}, \bibinfo{person}{Philippe Tillet}, \bibinfo{person}{Felipe~Petroski Such}, \bibinfo{person}{Dave Cummings}, \bibinfo{person}{Matthias Plappert}, \bibinfo{person}{Fotios
  Chantzis}, \bibinfo{person}{Elizabeth Barnes}, \bibinfo{person}{Ariel Herbert{-}Voss}, \bibinfo{person}{William~Hebgen Guss}, \bibinfo{person}{Alex Nichol}, \bibinfo{person}{Alex Paino}, \bibinfo{person}{Nikolas Tezak}, \bibinfo{person}{Jie Tang}, \bibinfo{person}{Igor Babuschkin}, \bibinfo{person}{Suchir Balaji}, \bibinfo{person}{Shantanu Jain}, \bibinfo{person}{William Saunders}, \bibinfo{person}{Christopher Hesse}, \bibinfo{person}{Andrew~N. Carr}, \bibinfo{person}{Jan Leike}, \bibinfo{person}{Joshua Achiam}, \bibinfo{person}{Vedant Misra}, \bibinfo{person}{Evan Morikawa}, \bibinfo{person}{Alec Radford}, \bibinfo{person}{Matthew Knight}, \bibinfo{person}{Miles Brundage}, \bibinfo{person}{Mira Murati}, \bibinfo{person}{Katie Mayer}, \bibinfo{person}{Peter Welinder}, \bibinfo{person}{Bob McGrew}, \bibinfo{person}{Dario Amodei}, \bibinfo{person}{Sam McCandlish}, \bibinfo{person}{Ilya Sutskever}, {and} \bibinfo{person}{Wojciech Zaremba}.} \bibinfo{year}{2021}\natexlab{}.
\newblock In \bibinfo{booktitle}{\emph{Evaluating Large Language Models Trained on Code}}, Vol.~\bibinfo{volume}{abs/2107.03374}. \bibinfo{publisher}{CoRR}.
\newblock
\showeprint[arXiv]{2107.03374}
\urldef\tempurl%
\url{https://arxiv.org/abs/2107.03374}
\showURL{%
\tempurl}


\bibitem[Chen et~al\mbox{.}(2024)]%
        {chen2024}
\bibfield{author}{\bibinfo{person}{Yinghao Chen}, \bibinfo{person}{Zehao Hu}, \bibinfo{person}{Chen Zhi}, \bibinfo{person}{Junxiao Han}, \bibinfo{person}{Shuiguang Deng}, {and} \bibinfo{person}{Jianwei Yin}.} \bibinfo{year}{2024}\natexlab{}.
\newblock \showarticletitle{ChatUniTest: A Framework for LLM-Based Test Generation}. In \bibinfo{booktitle}{\emph{Companion Proceedings of the 32nd ACM International Conference on the Foundations of Software Engineering}} (Porto de Galinhas, Brazil) \emph{(\bibinfo{series}{FSE 2024})}. \bibinfo{publisher}{Association for Computing Machinery}, \bibinfo{address}{New York, NY, USA}, \bibinfo{pages}{572–576}.
\newblock
\showISBNx{9798400706585}
\urldef\tempurl%
\url{https://doi.org/10.1145/3663529.3663801}
\showDOI{\tempurl}


\bibitem[Coignion et~al\mbox{.}(2024)]%
        {tristan2024}
\bibfield{author}{\bibinfo{person}{Tristan Coignion}, \bibinfo{person}{Cl\'{e}ment Quinton}, {and} \bibinfo{person}{Romain Rouvoy}.} \bibinfo{year}{2024}\natexlab{}.
\newblock \showarticletitle{A Performance Study of LLM-Generated Code on Leetcode}. In \bibinfo{booktitle}{\emph{Proceedings of the 28th International Conference on Evaluation and Assessment in Software Engineering}} (Salerno, Italy) \emph{(\bibinfo{series}{EASE '24})}. \bibinfo{publisher}{Association for Computing Machinery}, \bibinfo{address}{New York, NY, USA}, \bibinfo{pages}{79–89}.
\newblock
\showISBNx{9798400717017}
\urldef\tempurl%
\url{https://doi.org/10.1145/3661167.3661221}
\showDOI{\tempurl}


\bibitem[Daka and Fraser(2014)]%
        {daka2014}
\bibfield{author}{\bibinfo{person}{Ermira Daka} {and} \bibinfo{person}{Gordon Fraser}.} \bibinfo{year}{2014}\natexlab{}.
\newblock In \bibinfo{booktitle}{\emph{A Survey on Unit Testing Practices and Problems}}. \bibinfo{publisher}{2014 IEEE 25th International Symposium on Software Reliability Engineering}, \bibinfo{pages}{201--211}.
\newblock
\urldef\tempurl%
\url{https://doi.org/10.1109/ISSRE.2014.11}
\showDOI{\tempurl}


\bibitem[Damasceno et~al\mbox{.}(2023)]%
        {Damasceno_Bezerra_Campos_Machado_Coutinho_2023}
\bibfield{author}{\bibinfo{person}{Humberto Damasceno}, \bibinfo{person}{Carla Bezerra}, \bibinfo{person}{Denivan Campos}, \bibinfo{person}{Ivan Machado}, {and} \bibinfo{person}{Emanuel Coutinho}.} \bibinfo{year}{2023}\natexlab{}.
\newblock \showarticletitle{Test smell refactoring revisited: What can internal quality attributes and developers’ experience tell us?}
\newblock \bibinfo{journal}{\emph{Journal of Software Engineering Research and Development}} \bibinfo{volume}{11}, \bibinfo{number}{1} (\bibinfo{date}{Oct.} \bibinfo{year}{2023}), \bibinfo{pages}{13:1 – 13:16}.
\newblock
\urldef\tempurl%
\url{https://doi.org/10.5753/jserd.2023.3195}
\showDOI{\tempurl}


\bibitem[El~Haji et~al\mbox{.}(2024)]%
        {HAJI2024ACM}
\bibfield{author}{\bibinfo{person}{Khalid El~Haji}, \bibinfo{person}{Carolin Brandt}, {and} \bibinfo{person}{Andy Zaidman}.} \bibinfo{year}{2024}\natexlab{}.
\newblock In \bibinfo{booktitle}{\emph{Using {GitHub} Copilot for Test Generation in Python: An Empirical Study}} (Lisbon, Portugal) \emph{(\bibinfo{series}{AST '24})}. \bibinfo{publisher}{ACM}, \bibinfo{address}{New York, NY, USA}, \bibinfo{pages}{11}.
\newblock
\urldef\tempurl%
\url{https://doi.org/10.1145/3644032.3644443}
\showDOI{\tempurl}


\bibitem[Fakhoury et~al\mbox{.}(2024)]%
        {Fakhoury:TSE:2024}
\bibfield{author}{\bibinfo{person}{Sarah Fakhoury}, \bibinfo{person}{Aaditya Naik}, \bibinfo{person}{Georgios Sakkas}, \bibinfo{person}{Saikat Chakraborty}, {and} \bibinfo{person}{Shuvendu~K. Lahiri}.} \bibinfo{year}{2024}\natexlab{}.
\newblock \showarticletitle{LLM-Based Test-Driven Interactive Code Generation: User Study and Empirical Evaluation}.
\newblock \bibinfo{journal}{\emph{IEEE Transactions on Software Engineering}} \bibinfo{volume}{50}, \bibinfo{number}{9} (\bibinfo{year}{2024}), \bibinfo{pages}{2254--2268}.
\newblock
\urldef\tempurl%
\url{https://doi.org/10.1109/TSE.2024.3428972}
\showDOI{\tempurl}


\bibitem[Fernandes et~al\mbox{.}(2022)]%
        {tempy2022}
\bibfield{author}{\bibinfo{person}{Daniel Fernandes}, \bibinfo{person}{Ivan Machado}, {and} \bibinfo{person}{Rita Maciel}.} \bibinfo{year}{2022}\natexlab{}.
\newblock \showarticletitle{TEMPY: Test Smell Detector for Python}. In \bibinfo{booktitle}{\emph{Proceedings of the XXXVI Brazilian Symposium on Software Engineering}} (Virtual Event, Brazil) \emph{(\bibinfo{series}{SBES '22})}. \bibinfo{publisher}{Association for Computing Machinery}, \bibinfo{address}{New York, NY, USA}, \bibinfo{pages}{214–219}.
\newblock
\showISBNx{9781450397353}
\urldef\tempurl%
\url{https://doi.org/10.1145/3555228.3555280}
\showDOI{\tempurl}


\bibitem[Gonzalez et~al\mbox{.}(2017)]%
        {gonzales2017}
\bibfield{author}{\bibinfo{person}{Danielle Gonzalez}, \bibinfo{person}{Joanna~C.S. Santos}, \bibinfo{person}{Andrew Popovich}, \bibinfo{person}{Mehdi Mirakhorli}, {and} \bibinfo{person}{Mei Nagappan}.} \bibinfo{year}{2017}\natexlab{}.
\newblock \showarticletitle{A Large-Scale Study on the Usage of Testing Patterns That Address Maintainability Attributes: Patterns for Ease of Modification, Diagnoses, and Comprehension}. In \bibinfo{booktitle}{\emph{2017 IEEE/ACM 14th International Conference on Mining Software Repositories (MSR)}}. \bibinfo{pages}{391--401}.
\newblock
\urldef\tempurl%
\url{https://doi.org/10.1109/MSR.2017.8}
\showDOI{\tempurl}


\bibitem[Graham et~al\mbox{.}(2021)]%
        {graham2021foundations}
\bibfield{author}{\bibinfo{person}{D. Graham}, \bibinfo{person}{R. Black}, {and} \bibinfo{person}{E. van Veenendaal}.} \bibinfo{year}{2021}\natexlab{}.
\newblock \bibinfo{booktitle}{\emph{Foundations of Software Testing ISTQB Certification, 4th edition}}.
\newblock \bibinfo{publisher}{Cengage Learning}.
\newblock
\showISBNx{9780357884157}
\urldef\tempurl%
\url{https://books.google.com.br/books?id=mOwxEAAAQBAJ}
\showURL{%
\tempurl}


\bibitem[Hansson and Ellréus(2023)]%
        {Hansson_Ellréus_2023}
\bibfield{author}{\bibinfo{person}{Emilia Hansson} {and} \bibinfo{person}{Oliwer Ellréus}.} \bibinfo{year}{2023}\natexlab{}.
\newblock \emph{\bibinfo{title}{Code Correctness and Quality in the Era of AI Code Generation: Examining ChatGPT and GitHub Copilot}}.
\newblock \bibinfo{thesistype}{Ph.\,D. Dissertation}.
\newblock
\urldef\tempurl%
\url{https://urn.kb.se/resolve?urn=urn:nbn:se:lnu:diva-121545}
\showURL{%
\tempurl}


\bibitem[Khorikov(2020)]%
        {khorikov2020unit}
\bibfield{author}{\bibinfo{person}{V. Khorikov}.} \bibinfo{year}{2020}\natexlab{}.
\newblock \bibinfo{booktitle}{\emph{Unit Testing Principles, Practices, and Patterns: Effective testing styles, patterns, and reliable automation for unit testing, mocking, and integration testing with examples in C\#}}.
\newblock \bibinfo{publisher}{Manning}.
\newblock
\showISBNx{9781617296277}
\showLCCN{2020430603}
\urldef\tempurl%
\url{https://books.google.com.br/books?id=CbvZyAEACAAJ}
\showURL{%
\tempurl}


\bibitem[Kim(2020)]%
        {kim2020}
\bibfield{author}{\bibinfo{person}{Dong~Jae Kim}.} \bibinfo{year}{2020}\natexlab{}.
\newblock \showarticletitle{An Empirical Study on the Evolution of Test Smell}. In \bibinfo{booktitle}{\emph{2020 IEEE/ACM 42nd International Conference on Software Engineering: Companion Proceedings (ICSE-Companion)}}. \bibinfo{pages}{149--151}.
\newblock


\bibitem[Li(2022)]%
        {CHUN2022IEEE}
\bibfield{author}{\bibinfo{person}{Chun Li}.} \bibinfo{year}{2022}\natexlab{}.
\newblock In \bibinfo{booktitle}{\emph{Mobile GUI test script generation from natural language descriptions using pre-trained model}} (Pittsburgh, Pennsylvania) \emph{(\bibinfo{series}{MOBILESoft '22})}. \bibinfo{publisher}{Association for Computing Machinery}, \bibinfo{address}{New York, NY, USA}, \bibinfo{pages}{112–113}.
\newblock
\showISBNx{9781450393010}
\urldef\tempurl%
\url{https://doi.org/10.1145/3524613.3527809}
\showDOI{\tempurl}


\bibitem[Liu et~al\mbox{.}(2024)]%
        {liu2024}
\bibfield{author}{\bibinfo{person}{Chao Liu}, \bibinfo{person}{Xuanlin Bao}, \bibinfo{person}{Hongyu Zhang}, \bibinfo{person}{Neng Zhang}, \bibinfo{person}{Haibo Hu}, \bibinfo{person}{Xiaohong Zhang}, {and} \bibinfo{person}{Meng Yan}.} \bibinfo{year}{2024}\natexlab{}.
\newblock \showarticletitle{Guiding ChatGPT for Better Code Generation: An Empirical Study}. In \bibinfo{booktitle}{\emph{2024 IEEE International Conference on Software Analysis, Evolution and Reengineering (SANER)}}. \bibinfo{pages}{102--113}.
\newblock
\urldef\tempurl%
\url{https://doi.org/10.1109/SANER60148.2024.00018}
\showDOI{\tempurl}


\bibitem[Lu et~al\mbox{.}(2021)]%
        {codexglue}
\bibfield{author}{\bibinfo{person}{Shuai Lu}, \bibinfo{person}{Daya Guo}, \bibinfo{person}{Shuo Ren}, \bibinfo{person}{Junjie Huang}, \bibinfo{person}{Alexey Svyatkovskiy}, \bibinfo{person}{Ambrosio Blanco}, \bibinfo{person}{Colin Clement}, \bibinfo{person}{Dawn Drain}, \bibinfo{person}{Daxin Jiang}, \bibinfo{person}{Duyu Tang}, \bibinfo{person}{Ge Li}, \bibinfo{person}{Lidong Zhou}, \bibinfo{person}{Linjun Shou}, \bibinfo{person}{Long Zhou}, \bibinfo{person}{Michele Tufano}, \bibinfo{person}{Ming Gong}, \bibinfo{person}{Ming Zhou}, \bibinfo{person}{Nan Duan}, \bibinfo{person}{Neel Sundaresan}, \bibinfo{person}{Shao~Kun Deng}, \bibinfo{person}{Shengyu Fu}, {and} \bibinfo{person}{Shujie Liu}.} \bibinfo{year}{2021}\natexlab{}.
\newblock \bibinfo{title}{CodeXGLUE: A Machine Learning Benchmark Dataset for Code Understanding and Generation}.
\newblock
\newblock
\showeprint[arxiv]{2102.04664}~[cs.SE]
\urldef\tempurl%
\url{https://arxiv.org/abs/2102.04664}
\showURL{%
\tempurl}


\bibitem[Maragathavalli(2011)]%
        {Maragathavalli2011SearchbasedST}
\bibfield{author}{\bibinfo{person}{P. Maragathavalli}.} \bibinfo{year}{2011}\natexlab{}.
\newblock \showarticletitle{Search-based software test data generation using evolutionary computation}.
\newblock \bibinfo{journal}{\emph{ArXiv}}  \bibinfo{volume}{abs/1103.0125} (\bibinfo{year}{2011}).
\newblock
\urldef\tempurl%
\url{https://api.semanticscholar.org/CorpusID:25209645}
\showURL{%
\tempurl}


\bibitem[Marvin et~al\mbox{.}(2024)]%
        {Marvin2024}
\bibfield{author}{\bibinfo{person}{G. Marvin}, \bibinfo{person}{N. Hellen}, \bibinfo{person}{D. Jjingo}, {and} \bibinfo{person}{J. Nakatumba-Nabende}.} \bibinfo{year}{2024}\natexlab{}.
\newblock \showarticletitle{Engenharia de Prompt em Grandes Modelos de Linguagem}. In \bibinfo{booktitle}{\emph{Inteligência de Dados e Informática Cognitiva. ICDICI 2023}} \emph{(\bibinfo{series}{Algoritmos para Sistemas Inteligentes})}, \bibfield{editor}{\bibinfo{person}{IJ~Jacob}, \bibinfo{person}{S.~Piramuthu}, {and} \bibinfo{person}{P.~Falkowski-Gilski}} (Eds.). \bibinfo{publisher}{Springer}, \bibinfo{address}{Cingapura}.
\newblock
\urldef\tempurl%
\url{https://doi.org/10.1007/978-981-99-7962-2_30}
\showDOI{\tempurl}


\bibitem[Mathews and Nagappan(2024)]%
        {mathews2024}
\bibfield{author}{\bibinfo{person}{Noble~Saji Mathews} {and} \bibinfo{person}{Meiyappan Nagappan}.} \bibinfo{year}{2024}\natexlab{}.
\newblock \showarticletitle{Test-Driven Development and LLM-based Code Generation}. In \bibinfo{booktitle}{\emph{Proceedings of the 39th IEEE/ACM International Conference on Automated Software Engineering}} (Sacramento, CA, USA) \emph{(\bibinfo{series}{ASE '24})}. \bibinfo{publisher}{Association for Computing Machinery}, \bibinfo{address}{New York, NY, USA}, \bibinfo{pages}{1583–1594}.
\newblock
\showISBNx{9798400712487}
\urldef\tempurl%
\url{https://doi.org/10.1145/3691620.3695527}
\showDOI{\tempurl}


\bibitem[McMinn(2004)]%
        {mcminn2004}
\bibfield{author}{\bibinfo{person}{Phil McMinn}.} \bibinfo{year}{2004}\natexlab{}.
\newblock \showarticletitle{Search-based software test data generation: a survey: Research Articles}.
\newblock \bibinfo{journal}{\emph{Softw. Test. Verif. Reliab.}} \bibinfo{volume}{14}, \bibinfo{number}{2} (\bibinfo{date}{jun} \bibinfo{year}{2004}), \bibinfo{pages}{105–156}.
\newblock
\showISSN{0960-0833}


\bibitem[Oliveira et~al\mbox{.}(2024)]%
        {jhonatan2024}
\bibfield{author}{\bibinfo{person}{Jhonatan Oliveira}, \bibinfo{person}{Luigi Mateus}, \bibinfo{person}{Tássio Virgínio}, {and} \bibinfo{person}{Larissa Rocha}.} \bibinfo{year}{2024}\natexlab{}.
\newblock \showarticletitle{SNUTS.js: Sniffing Nasty Unit Test Smells in Javascript}. In \bibinfo{booktitle}{\emph{Anais do XXXVIII Simpósio Brasileiro de Engenharia de Software}} (Curitiba/PR). \bibinfo{publisher}{SBC}, \bibinfo{address}{Porto Alegre, RS, Brasil}, \bibinfo{pages}{720--726}.
\newblock
\showISSN{0000-0000}
\urldef\tempurl%
\url{https://doi.org/10.5753/sbes.2024.3563}
\showDOI{\tempurl}


\bibitem[Palomba et~al\mbox{.}(2018)]%
        {palomba2018}
\bibfield{author}{\bibinfo{person}{Fabio Palomba}, \bibinfo{person}{Andy Zaidman}, {and} \bibinfo{person}{Andrea De~Lucia}.} \bibinfo{year}{2018}\natexlab{}.
\newblock \showarticletitle{Automatic Test Smell Detection Using Information Retrieval Techniques}. In \bibinfo{booktitle}{\emph{2018 IEEE International Conference on Software Maintenance and Evolution (ICSME)}}. \bibinfo{pages}{311--322}.
\newblock
\urldef\tempurl%
\url{https://doi.org/10.1109/ICSME.2018.00040}
\showDOI{\tempurl}


\bibitem[Peng et~al\mbox{.}(2021)]%
        {PENG2021101347}
\bibfield{author}{\bibinfo{person}{Zedong Peng}, \bibinfo{person}{Xuanyi Lin}, \bibinfo{person}{Michelle Simon}, {and} \bibinfo{person}{Nan Niu}.} \bibinfo{year}{2021}\natexlab{}.
\newblock \showarticletitle{Unit and regression tests of scientific software: A study on SWMM}.
\newblock \bibinfo{journal}{\emph{Journal of Computational Science}}  \bibinfo{volume}{53} (\bibinfo{year}{2021}), \bibinfo{pages}{101347}.
\newblock
\showISSN{1877-7503}
\urldef\tempurl%
\url{https://doi.org/10.1016/j.jocs.2021.101347}
\showDOI{\tempurl}


\bibitem[Reeves et~al\mbox{.}(2023)]%
        {reeves2023}
\bibfield{author}{\bibinfo{person}{Brent Reeves}, \bibinfo{person}{Sami Sarsa}, \bibinfo{person}{James Prather}, \bibinfo{person}{Paul Denny}, \bibinfo{person}{Brett~A. Becker}, \bibinfo{person}{Arto Hellas}, \bibinfo{person}{Bailey Kimmel}, \bibinfo{person}{Garrett Powell}, {and} \bibinfo{person}{Juho Leinonen}.} \bibinfo{year}{2023}\natexlab{}.
\newblock \showarticletitle{Evaluating the Performance of Code Generation Models for Solving Parsons Problems With Small Prompt Variations}. In \bibinfo{booktitle}{\emph{Proceedings of the 2023 Conference on Innovation and Technology in Computer Science Education V. 1}} (Turku, Finland) \emph{(\bibinfo{series}{ITiCSE 2023})}. \bibinfo{publisher}{Association for Computing Machinery}, \bibinfo{address}{New York, NY, USA}, \bibinfo{pages}{299–305}.
\newblock
\showISBNx{9798400701382}
\urldef\tempurl%
\url{https://doi.org/10.1145/3587102.3588805}
\showDOI{\tempurl}


\bibitem[Ryan et~al\mbox{.}(2024)]%
        {ryan2024}
\bibfield{author}{\bibinfo{person}{Gabriel Ryan}, \bibinfo{person}{Siddhartha Jain}, \bibinfo{person}{Mingyue Shang}, \bibinfo{person}{Shiqi Wang}, \bibinfo{person}{Xiaofei Ma}, \bibinfo{person}{Murali~Krishna Ramanathan}, {and} \bibinfo{person}{Baishakhi Ray}.} \bibinfo{year}{2024}\natexlab{}.
\newblock \showarticletitle{Code-Aware Prompting: A Study of Coverage-Guided Test Generation in Regression Setting using LLM}.
\newblock \bibinfo{journal}{\emph{Proc. ACM Softw. Eng.}} \bibinfo{volume}{1}, \bibinfo{number}{FSE}, Article \bibinfo{articleno}{43} (\bibinfo{date}{July} \bibinfo{year}{2024}), \bibinfo{numpages}{21}~pages.
\newblock
\urldef\tempurl%
\url{https://doi.org/10.1145/3643769}
\showDOI{\tempurl}


\bibitem[Sahoo et~al\mbox{.}(2024)]%
        {sahoo2024systematicsurveypromptengineering}
\bibfield{author}{\bibinfo{person}{Pranab Sahoo}, \bibinfo{person}{Ayush~Kumar Singh}, \bibinfo{person}{Sriparna Saha}, \bibinfo{person}{Vinija Jain}, \bibinfo{person}{Samrat Mondal}, {and} \bibinfo{person}{Aman Chadha}.} \bibinfo{year}{2024}\natexlab{}.
\newblock \bibinfo{title}{A Systematic Survey of Prompt Engineering in Large Language Models: Techniques and Applications}.
\newblock
\newblock
\showeprint[arxiv]{2402.07927}~[cs.AI]
\urldef\tempurl%
\url{https://arxiv.org/abs/2402.07927}
\showURL{%
\tempurl}


\bibitem[Santana et~al\mbox{.}(2020)]%
        {santana2020raide}
\bibfield{author}{\bibinfo{person}{Railana Santana}, \bibinfo{person}{Luana Martins}, \bibinfo{person}{Larissa Rocha}, \bibinfo{person}{T{\'a}ssio Virg{\'\i}nio}, \bibinfo{person}{Adriana Cruz}, \bibinfo{person}{Heitor Costa}, {and} \bibinfo{person}{Ivan Machado}.} \bibinfo{year}{2020}\natexlab{}.
\newblock \showarticletitle{RAIDE: a tool for Assertion Roulette and Duplicate Assert identification and refactoring}. In \bibinfo{booktitle}{\emph{Proceedings of the XXXIV Brazilian Symposium on Software Engineering}}. \bibinfo{pages}{374--379}.
\newblock


\bibitem[Schäfer et~al\mbox{.}(2024)]%
        {SCHAFER2023IEEE}
\bibfield{author}{\bibinfo{person}{Max Schäfer}, \bibinfo{person}{Sarah Nadi}, \bibinfo{person}{Aryaz Eghbali}, {and} \bibinfo{person}{Frank Tip}.} \bibinfo{year}{2024}\natexlab{}.
\newblock In \bibinfo{booktitle}{\emph{An Empirical Evaluation of Using Large Language Models for Automated Unit Test Generation}}, Vol.~\bibinfo{volume}{50}. \bibinfo{publisher}{IEEE Transactions on Software Engineering}, \bibinfo{pages}{85--105}.
\newblock
\urldef\tempurl%
\url{https://doi.org/10.1109/TSE.2023.3334955}
\showDOI{\tempurl}


\bibitem[Shamshiri et~al\mbox{.}(2018)]%
        {shamshiri2018}
\bibfield{author}{\bibinfo{person}{Sina Shamshiri}, \bibinfo{person}{José~Miguel Rojas}, \bibinfo{person}{Juan~Pablo Galeotti}, \bibinfo{person}{Neil Walkinshaw}, {and} \bibinfo{person}{Gordon Fraser}.} \bibinfo{year}{2018}\natexlab{}.
\newblock \showarticletitle{How Do Automatically Generated Unit Tests Influence Software Maintenance?}. In \bibinfo{booktitle}{\emph{2018 IEEE 11th International Conference on Software Testing, Verification and Validation (ICST)}}. \bibinfo{pages}{250--261}.
\newblock
\urldef\tempurl%
\url{https://doi.org/10.1109/ICST.2018.00033}
\showDOI{\tempurl}


\bibitem[Siddiq et~al\mbox{.}(2024)]%
        {siddiq2024using}
\bibfield{author}{\bibinfo{person}{Mohammed~Latif Siddiq}, \bibinfo{person}{Joanna~Cecilia Da~Silva~Santos}, \bibinfo{person}{Ridwanul~Hasan Tanvir}, \bibinfo{person}{Noshin Ulfat}, \bibinfo{person}{Fahmid Al~Rifat}, {and} \bibinfo{person}{Vin\'{\i}cius Carvalho~Lopes}.} \bibinfo{year}{2024}\natexlab{}.
\newblock \showarticletitle{Using Large Language Models to Generate JUnit Tests: An Empirical Study}. In \bibinfo{booktitle}{\emph{Proceedings of the 28th International Conference on Evaluation and Assessment in Software Engineering}} (Salerno, Italy) \emph{(\bibinfo{series}{EASE '24})}. \bibinfo{publisher}{Association for Computing Machinery}, \bibinfo{address}{New York, NY, USA}, \bibinfo{pages}{313–322}.
\newblock
\showISBNx{9798400717017}
\urldef\tempurl%
\url{https://doi.org/10.1145/3661167.3661216}
\showDOI{\tempurl}


\bibitem[Tufano et~al\mbox{.}(2021)]%
        {tufano2021unit}
\bibfield{author}{\bibinfo{person}{Michele Tufano}, \bibinfo{person}{Dawn Drain}, \bibinfo{person}{Alexey Svyatkovskiy}, \bibinfo{person}{Shao~Kun Deng}, {and} \bibinfo{person}{Neel Sundaresan}.} \bibinfo{year}{2021}\natexlab{}.
\newblock \bibinfo{title}{Unit Test Case Generation with Transformers and Focal Context}.
\newblock
\newblock
\showeprint[arxiv]{2009.05617}~[cs.SE]


\bibitem[Tufano et~al\mbox{.}(2016)]%
        {tufano2016}
\bibfield{author}{\bibinfo{person}{Michele Tufano}, \bibinfo{person}{Fabio Palomba}, \bibinfo{person}{Gabriele Bavota}, \bibinfo{person}{Massimiliano Di~Penta}, \bibinfo{person}{Rocco Oliveto}, \bibinfo{person}{Andrea De~Lucia}, {and} \bibinfo{person}{Denys Poshyvanyk}.} \bibinfo{year}{2016}\natexlab{}.
\newblock \showarticletitle{An empirical investigation into the nature of test smells}. In \bibinfo{booktitle}{\emph{Proceedings of the 31st IEEE/ACM International Conference on Automated Software Engineering}} (Singapore, Singapore) \emph{(\bibinfo{series}{ASE '16})}. \bibinfo{publisher}{Association for Computing Machinery}, \bibinfo{address}{New York, NY, USA}, \bibinfo{pages}{4–15}.
\newblock
\showISBNx{9781450338455}
\urldef\tempurl%
\url{https://doi.org/10.1145/2970276.2970340}
\showDOI{\tempurl}


\bibitem[van Deursen et~al\mbox{.}(2001)]%
        {deursen2001}
\bibfield{author}{\bibinfo{person}{Arie van Deursen}, \bibinfo{person}{Leon Moonen}, \bibinfo{person}{Alex van~den Bergh}, {and} \bibinfo{person}{Gerard Kok}.} \bibinfo{year}{2001}\natexlab{}.
\newblock In \bibinfo{booktitle}{\emph{Refactoring Test Code}}, \bibfield{editor}{\bibinfo{person}{M.~Marchesi} {and} \bibinfo{person}{G.~Succi}} (Eds.). \bibinfo{publisher}{Proceedings 2nd International Conference on Extreme Programming and Flexible Processes in Software Engineering (XP2001)}.
\newblock


\bibitem[Wang et~al\mbox{.}(2022)]%
        {pynose}
\bibfield{author}{\bibinfo{person}{Tongjie Wang}, \bibinfo{person}{Yaroslav Golubev}, \bibinfo{person}{Oleg Smirnov}, \bibinfo{person}{Jiawei Li}, \bibinfo{person}{Timofey Bryksin}, {and} \bibinfo{person}{Iftekhar Ahmed}.} \bibinfo{year}{2022}\natexlab{}.
\newblock In \bibinfo{booktitle}{\emph{PyNose: a test smell detector for python}} (Melbourne, Australia) \emph{(\bibinfo{series}{ASE '21})}. \bibinfo{publisher}{IEEE Press}, \bibinfo{pages}{593–605}.
\newblock
\showISBNx{9781665403375}
\urldef\tempurl%
\url{https://doi.org/10.1109/ASE51524.2021.9678615}
\showDOI{\tempurl}


\bibitem[Wang et~al\mbox{.}(2024)]%
        {wang2024}
\bibfield{author}{\bibinfo{person}{Zejun Wang}, \bibinfo{person}{Kaibo Liu}, \bibinfo{person}{Ge Li}, {and} \bibinfo{person}{Zhi Jin}.} \bibinfo{year}{2024}\natexlab{}.
\newblock \showarticletitle{HITS: High-coverage LLM-based Unit Test Generation via Method Slicing}. In \bibinfo{booktitle}{\emph{Proceedings of the 39th IEEE/ACM International Conference on Automated Software Engineering}} (Sacramento, CA, USA) \emph{(\bibinfo{series}{ASE '24})}. \bibinfo{publisher}{Association for Computing Machinery}, \bibinfo{address}{New York, NY, USA}, \bibinfo{pages}{1258–1268}.
\newblock
\showISBNx{9798400712487}
\urldef\tempurl%
\url{https://doi.org/10.1145/3691620.3695501}
\showDOI{\tempurl}


\bibitem[Xie and Notkin(2006)]%
        {XieN2006}
\bibfield{author}{\bibinfo{person}{Tao Xie} {and} \bibinfo{person}{David Notkin}.} \bibinfo{year}{2006}\natexlab{}.
\newblock \showarticletitle{Tool-assisted unit test generation and selection based on operational abstractions}.
\newblock \bibinfo{journal}{\emph{Automated Software Engineering Journal}} \bibinfo{volume}{13}, \bibinfo{number}{3} (\bibinfo{date}{July} \bibinfo{year}{2006}), \bibinfo{pages}{345--371}.
\newblock


\bibitem[Yetistiren et~al\mbox{.}(2022)]%
        {yetistiren2022acm}
\bibfield{author}{\bibinfo{person}{Burak Yetistiren}, \bibinfo{person}{Isik Ozsoy}, {and} \bibinfo{person}{Eray Tuzun}.} \bibinfo{year}{2022}\natexlab{}.
\newblock \showarticletitle{Assessing the quality of GitHub copilot’s code generation}. In \bibinfo{booktitle}{\emph{Assessing the quality of GitHub copilot’s code generation}} (Singapore, Singapore) \emph{(\bibinfo{series}{PROMISE 2022})}. \bibinfo{publisher}{Association for Computing Machinery}, \bibinfo{address}{New York, NY, USA}, \bibinfo{pages}{62–71}.
\newblock
\showISBNx{9781450398602}
\urldef\tempurl%
\url{https://doi.org/10.1145/3558489.3559072}
\showDOI{\tempurl}


\bibitem[Yetiştiren et~al\mbox{.}(2023)]%
        {yetistiren2023}
\bibfield{author}{\bibinfo{person}{Burak Yetiştiren}, \bibinfo{person}{Işık Özsoy}, \bibinfo{person}{Miray Ayerdem}, {and} \bibinfo{person}{Eray Tüzün}.} \bibinfo{year}{2023}\natexlab{}.
\newblock In \bibinfo{booktitle}{\emph{Evaluating the Code Quality of AI-Assisted Code Generation Tools: An Empirical Study on GitHub Copilot, Amazon CodeWhisperer, and ChatGPT}}.
\newblock
\urldef\tempurl%
\url{https://doi.org/10.48550/arXiv.2304.10778}
\showDOI{\tempurl}


\bibitem[Yu et~al\mbox{.}(2023)]%
        {shengcheng2023ieee}
\bibfield{author}{\bibinfo{person}{Shengcheng Yu}, \bibinfo{person}{Chunrong Fang}, \bibinfo{person}{Yucheng Ling}, \bibinfo{person}{Chentian Wu}, {and} \bibinfo{person}{Zhenyu Chen}.} \bibinfo{year}{2023}\natexlab{}.
\newblock \showarticletitle{LLM for Test Script Generation and Migration: Challenges, Capabilities, and Opportunities}. In \bibinfo{booktitle}{\emph{LLM for Test Script Generation and Migration: Challenges, Capabilities, and Opportunities}}. \bibinfo{publisher}{2023 IEEE 23rd International Conference on Software Quality, Reliability, and Security (QRS)}, \bibinfo{pages}{206--217}.
\newblock
\urldef\tempurl%
\url{https://api.semanticscholar.org/CorpusID:262459296}
\showURL{%
\tempurl}


\end{thebibliography}

\end{document}